\documentclass[12pt]{article}

\usepackage{cite}
\usepackage{epsfig}
\usepackage{amsmath}
\usepackage{pstricks}

\def\mathswitch#1{\relax\ifmmode#1\else$#1$\fi}
\def\mathswitchr#1{\relax\ifmmode{\mathrm{#1}}\else$\mathrm{#1}$\fi}

\newcommand{\SLASH}[2]{\makebox[#2ex][l]{$#1$}/}

\newcommand{\anc}{\rule{0mm}{0mm}}

\newcommand{\knickpfeil}{\;\raisebox{1.12ex}{$\lfloor$} \!\!\! \to}

\newcommand{\mycaption}[1]{\caption{\sl #1}}

\begin{document}
\thispagestyle{empty}

\def\thefootnote{\fnsymbol{footnote}}

\begin{flushright}
\end{flushright}

\vspace{1cm}

\begin{center}

{\Large\sc {\bf General analysis of decay chains with three-body decays involving
missing energy}}
\\[3.5em]
{\large\sc
Chien-Yi~Chen$^1$, A.~Freitas$^2$
}

\vspace*{1cm}

{\sl $^1$ Department of Physics, Carnegie Mellon University, Pittsburgh, PA
15213, USA
\\[1em]
\sl $^2$
Pittsburgh Particle-physics Astro-physics \& Cosmology Center
(Pitt-PACC),\\ Department of Physics \& Astronomy, University of Pittsburgh,
Pittsburgh, PA 15260, USA
}

\end{center}

\vspace*{2.5cm}

\begin{abstract}

A model-independent analysis of decays of the form $C \to \ell^+\ell^-A$
($\ell=e,\mu$) is presented, including the possibility that this three-body
decay is preceded by an additional decay step $D \to j C$. Here $A$, $C$ and $D$
are heavy new-physics particles and $j$ stands for a quark jet.  It is assumed
that $A$ escapes direct detection in a collider experiment, so that one cannot
kinematically reconstruct the momenta of the new particles. Instead, information
about their properties can be obtained from invariant-mass distributions of the
visible decay products, $i.\,e.$ the di-lepton ($\ell\ell$) and jet-lepton
($j\ell$) invariant-mass distributions.
All possible spin configurations and renormalizable couplings of the new
particles are considered, and explicit expressions for the invariant-mass
distributions are derived, in a formulation that separates the coupling
parameters from the spin and kinematic information. In a numerical
analysis, it is shown how these properties can be determined independently from
a fit to the $m_{\ell\ell}$ and $m_{j\ell}$ distributions.

\end{abstract}

\setcounter{page}{0}
\setcounter{footnote}{0}

\newpage


\section{Introduction}

A large range of models have been proposed that predict new particles within the
reach of the Large Hadron Collider (LHC). Since there is currently very little
evidence for favoring one model over the others, it will be essential to analyze
a potential new-physics signal in the LHC data in a model-independent approach,
by independently determining the properties of each of the produced particles.
Recently, this idea has gained increased interest, and several groups have
worked on constructing such model-independent setups for a number of different
observable signatures, see $e.\,g.$
Refs.~\cite{modelind,modelindmet,spin2,Burns:2008cp}. A particularly challenging
scenario are processes that result in the production of new weakly interacting massive
particles (WIMPs), which are invisible to the detector. WIMPs are predicted in
many models as hypothetical dark matter candidates. 
In these models, the stability of the WIMP
is a consequence of some (discrete) symmetry, under which it is charged.  As a
result, it can be produced only in pairs at colliders, leading to challenging
events with at least two invisible objects.
At hadron colliders like the LHC there are not enough kinematical constraints in
events of this type for the direct reconstruction of the momenta of all
particles involved.

One approach to this problem is motivated by the fact that models predict
additional new particles, which can decay into the stable WIMP. In this case, one
can have cascade decay chains, which go through multiple decay steps before
ending with the stable WIMP, so that one can  construct invariant-mass
distributions of the visible decay products of this cascade. The kinematic
endpoints of these distributions yield information about the masses \cite{mass} 
of the new heavy particles, while  the shape is sensitive to their spins
\cite{spin,spin2,Miller:2005zp,Burns:2008cp}. Refs.~\cite{spin2,Burns:2008cp}
have analyzed decay chains built up from a sequence of two-body decays in a
model-independent way, by considering arbitrary spin assignments \cite{spin2}
and also using general parametrizations for the coupling for the new particles
\cite{Burns:2008cp}.

However, for scenarios with relatively small splittings in the mass spectrum of
the new-physics particles, it can often happen that the last decay step is a
three-body decay mediated by a heavier off-shell particle, see right-hand side
of Fig.~\ref{fig:diag}. In Ref.~\cite{Csaki:2007xm}, three-body decays have been
analyzed in order to distinguish gluinos, the supersymmetric partners of gluons,
from a Kaluza-Klein (KK) gluons in universal extra dimensions (UED). A
model-independent study of three-body decays has been presented in
Ref.~\cite{Edelhauser:2010gb}, but only in the limit of an asymptotically large
mass of the intermediate off-shell particles. In typical supersymmetry and UED
scenarios, however, this limit is often not a good approximation.

\begin{figure}[t]
\anc\\[5cm]
\psline[linewidth=1.5pt](0,2)(6,2)
\psline[linewidth=1.5pt](2,2)(2.5,3.5)
\psline[linewidth=1.5pt](4,2)(4.5,3.5)
\psline[linewidth=1.5pt](4,2)(5.1,3.2)
\psdot[dotsize=5pt](2,2)
\pscircle[fillstyle=solid,fillcolor=gray](4,2){.3}
\rput[b](1,2.2){$D$}
\rput[b](3.2,2.2){$C$}
\rput[b](5.3,2.2){$A$}
\rput[r](2.3,3.5){$q$}
\rput[r](4.3,3.5){$\ell^-$}
\rput[l](5.3,3.2){$\ell^+$}
\psframe[linestyle=dashed](2.7,1.3)(6.5,4)
\psline[linestyle=dotted](6.5,4)(8,5.5)
\psline[linestyle=dotted](6.5,1.3)(8,-0.3)
\psframe[linestyle=dashed](8,-0.3)(16.5,5.5)
\rput[l](8.2,4.8){Topology I:}
\psline[arrowsize=4pt 3]{-<}(9,3.5)(10.7,3.5)
\psline[linewidth=1.5pt](9,3.5)(12,3.5)
\psline[arrowsize=4pt 3]{->}(10,3.5)(10.35,4.2)
\psline[linewidth=1.5pt](10,3.5)(10.5,4.5)
\psline[arrowsize=4pt 3]{<-}(11.15,3.8)(11.5,4.5)
\psline[linewidth=1.5pt](11,3.5)(11.5,4.5)
\psdot[dotsize=5pt](10,3.5)
\psdot[dotsize=5pt](11,3.5)
\rput[t](9.5,3.3){$C$}
\rput[t](10.5,3.3){$B^*$}
\rput[t](11.5,3.3){$A$}
\rput[l](10.6,4.5){$\ell^-$}
\rput[l](11.6,4.5){$\ell^+$}
\psline[arrowsize=4pt 3]{->}(13,3.5)(14.7,3.5)
\psline[linewidth=1.5pt](13,3.5)(16,3.5)
\psline[arrowsize=4pt 3]{-<}(14,3.5)(15.2,4.3)
\psline[linewidth=1.5pt](14,3.5)(15.5,4.5)
\psline[arrowsize=4pt 3]{>-}(14.7,4.1)(14.5,4.5)
\psline[linewidth=1.5pt](15,3.5)(14.5,4.5)
\psdot[dotsize=5pt](14,3.5)
\psdot[dotsize=5pt](15,3.5)
\rput[t](13.5,3.3){$C$}
\rput[t](14.5,3.3){$B^*$}
\rput[t](15.5,3.3){$A$}
\rput[r](14.4,4.5){$\ell^-$}
\rput[l](15.6,4.5){$\ell^+$}
\rput(12.5,4){\large\bf +}
\rput[l](8.2,2){Topology II:}
\psline[linewidth=1.5pt](9,0.5)(12,0.5)
\rput{63.4}(10.55,0.55){\pscurve[linewidth=1.5pt]%
	(0,0)(0.1,0.05)(0.2,0)(0.3,-0.05)(0.4,0)(0.5,0.05)%
	(0.6,0)(0.7,-0.05)(0.8,0)(0.9,0.05)(1,0)}
\psline[linewidth=1.5pt](12,2)(11,1.5)(12,1)
\psdot[dotsize=5pt](10.5,0.5)
\psdot[dotsize=5pt](11,1.5)
\rput[t](9.5,0.3){$C$}
\rput[t](11.5,0.3){$A$}
\rput[r](10.5,1){$Z^*$}
\rput[l](12.1,2){$\ell^-$}
\rput[l](12.1,1.1){$\ell^+$}
\vspace{1ex}
\mycaption{Right: Three-body decays involving an off-shell new-physics particle
B (topology~I) or an off-shell Z boson (topology~II). Left: The three body decay
could occur as the last step of a longer decay chain.}
\label{fig:diag}
\end{figure}

In this work, three-body decays of the form $C \to \ell^+\ell^- A$ will be
analyzed in a model-independent setup without assumptions about the values of
the masses of the new-physics particles. Here $C$ is a massive new particle that
decays into the WIMP $A$ and two SM leptons $\ell^\pm=e^\pm,\mu^\pm$ through the
off-shell exchange of a third new particle $B$ or the SM $Z$-boson, see
Fig.~\ref{fig:diag}\footnote{In general, besides the $Z$-boson, a bosonic
new-physics particle ($e.\,g.$ a $Z'$ or a Higgs boson) may also appear in
the decay topology II. However, the branching of such a particle into leptons
is strongly constrained by data on four-fermion contact interactions
\cite{lep4l}, and thus its contribution will be neglected here.}. The spins of
$A$, $B$, and $C$, their coupling parameters, and the mass $m_B$ of the particle
$B$ will be kept as free quantities that have to be extracted from the
experimental data. We only impose the constraint $m_B>m_C$, or $m_Z > m_C-m_A$,
to ensure that we have an actual three-body decay. Without these constraints the
three-body decay would decompose into two two-body decays, which is a scenario
that has been discussed in detail in the literature cited above.

Furthermore, we also consider the case that this three-body decay is the second
step of a cascade decay of the form $D \to 
\stackrel{\text{\tiny (--)}}{q}\! C \to
\stackrel{\text{\tiny (--)}}{q}\! \ell^+ \ell^- A$, where 
$\stackrel{\text{\tiny (--)}}{q}\!$ refers to a SM
quark (antiquark), see Fig.~\ref{fig:diag}.
Such a decay chain would lead to two independent observable invariant-mass
distributions, a di-lepton ($\ell^+\ell^-$) invariant-mass distribution, and a
jet-lepton ($j\ell^\pm$) invariant-mass distribution, where the jet emerges from
the fragmentation of the quark or antiquark.

For both of these cases, we investigate the simultaneous determination of the
spins and couplings of the new particles $A$, $B$, $C$ and $D$ from the shapes
of these distributions. The determination of the masses from kinematic endpoints
has been discussed elsewhere \cite{mass}, and here we will simply assume that
the masses of the particles $A$, $C$ and $D$ are already known. On the other
hand, the mass $m_B$ of the off-shell intermediate particle $B$ can not be
extracted from the kinematic endpoints, and we will study  if instead it can be
constrained from the shapes of the distributions.

Our analysis closely follows the conventions of Ref.~\cite{Burns:2008cp}. After
introducing the relevant spin and coupling representations in
section~\ref{setup}, the calculation of the $\ell\ell$ and $j\ell$
invariant-mass distributions is described in section~\ref{dist}. In
section~\ref{ana} we present a procedure for determining the spins and couplings
of the new particles, as well as the mass of the intermediate particle $B$, by
fitting the theoretically calculated functions to the experimentally observed
distributions. The method is illustrated by applying it in two numerical
examples.
Finally, our main conclusions are summarized in section~\ref{summ}.


\section{Setup}
\label{setup}

The three-body decay of a heavy new particle $C$ into two opposite-sign
same-flavor leptons and a second new particle $A$,
\begin{equation}
C \to \ell^+\ell^- A, \qquad (\ell = e,\mu), \label{3b}
\end{equation}
is mediated either by an off-shell heavy new particle $B$ (with $m_B > m_C >
m_A$) or a SM $Z$-boson (with $m_C - m_A > m_Z$).
We also consider the possibility that eq.~\eqref{3b} occurs as the last step of
a longer decay chain,
\begin{equation}
\begin{aligned}
D \to q \, &C \\[-1ex]
&\knickpfeil \ell^+\ell^- A.
\end{aligned} \label{chain}
\end{equation}
Here $D$ is a QCD triplet, while $B$ and $A/C$ are electrically charged and
neutral QCD singlets, respectively. For the purpose of this work, it is assumed
that $A$ and $C$ are self-conjugate ($i.\,e.$ they are their own antiparticles)%
\footnote{Some new physics models predict decay chains with non-self-conjugate
neutral heavy particles, which lead to distinct phenomenological features
\cite{majdir}, but this case will not be considered here.}.
Furthermore, it is assumed that $A$, $B$, $C$, and $D$ are charged under some symmetry which
ensures that $A$ is stable and escapes from the detector without leaving a
signal.

In general, it is difficult to experimentally determine the overall strength of
the couplings in the decay chain since the width of weakly decaying particles is
typically small compared to the experimental resolution. Consequently, only the
\emph{shape} of the observable invariant-mass distributions will be considered
here, similar to earlier studies on spin determination
\cite{spin,spin2,Burns:2008cp,Csaki:2007xm,Edelhauser:2010gb}. All expressions
for these distributions presented in the following sections
therefore include an arbitrary, but constant, normalization factor.

\begin{table}[t]
\renewcommand{\arraystretch}{1.2}
\anc\hspace{4cm}
\begin{tabular}{|r|cccc|l|}
\hline
$S$ & $D$ & $C$ & $B$ & $A$ & Example \\
\hline\hline
1 & S&F&S&F & $\tilde{q} \to \tilde{\chi}^0_2 \to \tilde{\ell}^* \to \tilde{\chi}^0_1$ \\
\hline
2 & F&S&F&S & $q_{(1)} \to W^0_{H,(1)} \to \ell_{(1)}^* \to B^0_{H,(1)}$ \\
\hline
3 & F&S&F&V & $q_{(1)} \to W^0_{H,(1)} \to \ell_{(1)}^* \to B^0_{\mu,(1)}$ \\
\hline
\rput[r](-1em,1.1em){Topology I\phantom{I} $\left\{\rule{0mm}{10.8ex}\right.$}
4 & F&V&F&S & $q_{(1)} \to W^0_{\mu,(1)} \to \ell_{(1)}^* \to B^0_{H,(1)}$ \\
\hline
5 & F&V&F&V & $q_{(1)} \to W^0_{\mu,(1)} \to \ell_{(1)}^* \to B^0_{\mu,(1)}$ \\
\hline
6 & S&F&V&F &  \\
\hline\hline
7 & F&S&&S & $q_{(1)} \to W^0_{H,(1)} \to B^0_{H,(1)}$ \\
\hline
8 & F&S&&V & $q_{(1)} \to W^0_{H,(1)} \to B^0_{\mu,(1)}$ \\
\hline
\rput[r](-1em,.2em){Topology II $\left\{\rule{0mm}{9ex}\right.$}
9 & F&V&&S & $q_{(1)} \to W^0_{\mu,(1)} \to B^0_{H,(1)}$ \\
\hline
10 & F&V&&V & $q_{(1)} \to W^0_{\mu,(1)} \to B^0_{\mu,(1)}$ \\
\hline
11 & S&F&&F & $\tilde{q} \to \tilde{\chi}^0_2 \to \tilde{\chi}^0_1$ \\
\hline
\end{tabular}
\mycaption{Possible spin configurations of the heavy particles D, C, B, and A in
the decay chain of Fig.~\ref{fig:diag} (F=Fermion, S=Scalar, V=Vector). 
Also shown are examples for realizations of these assignments
in the Minimal Supersymmetric Standard Model (MSSM) or in models with one or two
universal extra dimension (UED). Here $\tilde{q}$, $\tilde{\ell}$,
and $\tilde{\chi}^0_i$ denote squark, slepton, and neutralino, respectively.
$q_{(1)}$, $\ell_{(1)}$, $\tilde{B}^0_{\mu,(1)}$, and
$\tilde{W}^{0,\pm}_{\mu,(1)}$ refer to the
first-level KK-excitations of quark, lepton, U(1) gauge field, and SU(2) gauge
field, respectively. $B_{H,(1)}^0$ and $W_{H,(1)}^0$ are
scalars stemming from one of the extra components of the higher-dimensional
gauge fields in UED. More details of these models can be found in
Refs.~\cite{susy,ued}.}
\label{tab:spins}
\end{table}

Table~\ref{tab:spins} lists all possible spin assignments for the particles
$A{-}D$ in any renormalizable theory with fields of spin 0 (scalars), spin 1/2
(fermions) and/or spin 1 (vector bosons). Also shown are examples for
realizations of these assignments in known models.

The chirality of the fermion couplings depend on the details of the new physics
and thus are a priori unknown. Following Ref.~\cite{Burns:2008cp}, we
introduce arbitrary left- and right-handed components. For
scalar-fermion-fermion vertices, the interaction Lagrangians are defined as
\begin{align}
&\parbox[c][1cm]{3cm}{
\psline[linewidth=1.5pt](0.5,0)(1.5,0)
\psline[linewidth=1.5pt](1.5,0)(2.5,0.3)
\psline[linewidth=1.5pt,linestyle=dashed](1.5,0)(2.5,-0.3)
\rput[r](0.4,0){$B$}
\rput[l](2.65,0.3){$\ell$}
\rput[l](2.6,-0.3){$A$}} \qquad
\overline{\psi}_B \, A \, (a_L \omega_- + a_R \omega_+) \, \psi_\ell + \text{h.c.},
\label{lagr1} \\[1ex]
&\parbox[c][1cm]{3cm}{
\psline[linewidth=1.5pt,linestyle=dashed](0.5,0)(1.5,0)
\psline[linewidth=1.5pt](1.5,0)(2.5,0.3)
\psline[linewidth=1.5pt](1.5,0)(2.5,-0.3)
\rput[r](0.4,0){$B$}
\rput[l](2.65,0.3){$\ell$}
\rput[l](2.6,-0.3){$A$}} \qquad
\overline{\psi}_A \, B \, (a_L \omega_- + a_R \omega_+) \, \psi_\ell + \text{h.c.},
\displaybreak[0] \\[1ex]
&\parbox[c][1cm]{3cm}{
\psline[linewidth=1.5pt](0.5,0)(1.5,0)
\psline[linewidth=1.5pt](1.5,0)(2.5,0.3)
\psline[linewidth=1.5pt,linestyle=dashed](1.5,0)(2.5,-0.3)
\rput[r](0.4,0){$C$}
\rput[l](2.65,0.3){$\ell$}
\rput[l](2.6,-0.3){$B$}} \qquad
\overline{\psi}_C \, B \, (b_L \omega_- + b_R \omega_+) \, \psi_\ell + \text{h.c.},
\\[1ex]
&\parbox[c][1cm]{3cm}{
\psline[linewidth=1.5pt,linestyle=dashed](0.5,0)(1.5,0)
\psline[linewidth=1.5pt](1.5,0)(2.5,0.3)
\psline[linewidth=1.5pt](1.5,0)(2.5,-0.3)
\rput[r](0.4,0){$C$}
\rput[l](2.65,0.3){$\ell$}
\rput[l](2.6,-0.3){$B$}} \qquad
\overline{\psi}_B \, C \, (b_L \omega_- + b_R \omega_+) \, \psi_\ell + \text{h.c.},
\displaybreak[0] \\[1ex]
&\parbox[c][1cm]{3cm}{
\psline[linewidth=1.5pt](0.5,0)(1.5,0)
\psline[linewidth=1.5pt](1.5,0)(2.5,0.3)
\psline[linewidth=1.5pt,linestyle=dashed](1.5,0)(2.5,-0.3)
\rput[r](0.4,0){$D$}
\rput[l](2.65,0.3){$q$}
\rput[l](2.6,-0.3){$C$}} \qquad
\overline{\psi}_D \, C \, (c_L \omega_- + c_R \omega_+) \, \psi_q + \text{h.c.},
\\[1ex]
&\parbox[c][1cm]{3cm}{
\psline[linewidth=1.5pt,linestyle=dashed](0.5,0)(1.5,0)
\psline[linewidth=1.5pt](1.5,0)(2.5,0.3)
\psline[linewidth=1.5pt](1.5,0)(2.5,-0.3)
\rput[r](0.4,0){$D$}
\rput[l](2.65,0.3){$q$}
\rput[l](2.6,-0.3){$C$}} \qquad
\overline{\psi}_C \, D \, (c_L \omega_- + c_R \omega_+) \, \psi_q + \text{h.c.},
\end{align}
where $\omega_\pm = \frac{1}{2}(1\pm\gamma_5)$. For vector-fermion-fermion
couplings, $A$ must be replaced by $\SLASH{A}{.5}\,$ in \eqref{lagr1},
\emph{etc}.
After normalizing the overall coupling strength to unity, each vertex can be
parametrized by a single angle $\alpha$, $\beta$, or $\gamma$,
\begin{equation}
\begin{aligned}
a_L &= \cos\alpha, \quad & 
b_L &= \cos\beta, \quad & 
c_L &= \cos\gamma, \\
a_R &= \sin\alpha, & 
b_R &= \sin\beta, & 
c_R &= \sin\gamma.
\end{aligned}
\end{equation}
As will be shown later,
the entire parameter space for the couplings can be
covered by restricting the angles to the intervals $\alpha \in
[-\pi/2,\pi/2],\,\beta,\gamma \in [0,\pi/2]$.

The form of the $CAZ$ vertices is uniquely determined by Lorentz symmetry and CP
properties (since the $Z$-boson is CP-odd, while the self-conjugate $A$ and $C$
are C-even):
\begin{align}
&\parbox[c][1cm]{3cm}{
\psline[linewidth=1.5pt,linestyle=dashed](0.5,0)(1.5,0)
\rput{20}(1.5,0){\pscurve[linewidth=1.5pt]%
	(0,0)(0.1,0.05)(0.2,0)(0.3,-0.05)(0.4,0)(0.5,0.05)%
	(0.6,0)(0.7,-0.05)(0.8,0)(0.9,0.05)(1,0)}
\rput{-20}(1.5,0){\psline[linewidth=1.5pt,linestyle=dashed](0,0)(1,0)}
\rput[r](0.4,0){$C$}
\rput[l](2.5,0.3){$Z$}
\rput[l](2.5,-0.3){$A$}} \qquad iC\!\! 
 \stackrel{\leftrightarrow}{\partial}_\mu \!A \, Z^\mu ,
\\[1ex]
&\parbox[c][1cm]{3cm}{
\rput(0.5,0){\pscurve[linewidth=1.5pt]%
	(0,0)(0.1,0.05)(0.2,0)(0.3,-0.05)(0.4,0)(0.5,0.05)%
	(0.6,0)(0.7,-0.05)(0.8,0)(0.9,0.05)(1,0)}
\rput{20}(1.5,0){\pscurve[linewidth=1.5pt]%
	(0,0)(0.1,0.05)(0.2,0)(0.3,-0.05)(0.4,0)(0.5,0.05)%
	(0.6,0)(0.7,-0.05)(0.8,0)(0.9,0.05)(1,0)}
\rput{-20}(1.5,0){\psline[linewidth=1.5pt,linestyle=dashed](0,0)(1,0)}
\rput[r](0.4,0){$C$}
\rput[l](2.5,0.3){$Z$}
\rput[l](2.5,-0.3){$A$}} \qquad -C_\mu \,A \, Z^\mu ,
\\[1ex]
&\parbox[c][1cm]{3cm}{
\psline[linewidth=1.5pt,linestyle=dashed](0.5,0)(1.5,0)
\rput{20}(1.5,0){\pscurve[linewidth=1.5pt]%
	(0,0)(0.1,0.05)(0.2,0)(0.3,-0.05)(0.4,0)(0.5,0.05)%
	(0.6,0)(0.7,-0.05)(0.8,0)(0.9,0.05)(1,0)}
\rput{-20}(1.5,0){\pscurve[linewidth=1.5pt]%
	(0,0)(0.1,0.05)(0.2,0)(0.3,-0.05)(0.4,0)(0.5,0.05)%
	(0.6,0)(0.7,-0.05)(0.8,0)(0.9,0.05)(1,0)}
\rput[r](0.4,0){$C$}
\rput[l](2.5,0.3){$Z$}
\rput[l](2.5,-0.3){$A$}} \qquad -C \,A_\mu \, Z^\mu,
\\[1ex]
&\parbox[c][1cm]{3cm}{
\rput(0.5,0){\pscurve[linewidth=1.5pt]%
	(0,0)(0.1,0.05)(0.2,0)(0.3,-0.05)(0.4,0)(0.5,0.05)%
	(0.6,0)(0.7,-0.05)(0.8,0)(0.9,0.05)(1,0)}
\rput{20}(1.5,0){\pscurve[linewidth=1.5pt]%
	(0,0)(0.1,0.05)(0.2,0)(0.3,-0.05)(0.4,0)(0.5,0.05)%
	(0.6,0)(0.7,-0.05)(0.8,0)(0.9,0.05)(1,0)}
\rput{-20}(1.5,0){\pscurve[linewidth=1.5pt]%
	(0,0)(0.1,0.05)(0.2,0)(0.3,-0.05)(0.4,0)(0.5,0.05)%
	(0.6,0)(0.7,-0.05)(0.8,0)(0.9,0.05)(1,0)}
\rput[r](0.4,0){$C$}
\rput[l](2.5,0.3){$Z$}
\rput[l](2.5,-0.3){$A$}} \qquad (C_\mu A_\nu - A_\mu C_\nu) \partial^\mu Z^\nu +
\text{cycl.},
\\[1ex]
&\parbox[c][1cm]{3cm}{
\psline[linewidth=1.5pt](0.5,0)(1.5,0)
\rput{20}(1.5,0){\pscurve[linewidth=1.5pt]%
	(0,0)(0.1,0.05)(0.2,0)(0.3,-0.05)(0.4,0)(0.5,0.05)%
	(0.6,0)(0.7,-0.05)(0.8,0)(0.9,0.05)(1,0)}
\rput{-20}(1.5,0){\psline[linewidth=1.5pt](0,0)(1,0)}
\rput[r](0.4,0){$C$}
\rput[l](2.5,0.3){$Z$}
\rput[l](2.5,-0.3){$A$}} \qquad \overline{\psi}_C
 \gamma_\mu \gamma_5 \psi_A \, Z^\mu ,
\end{align}
where again the coupling constants have been normalized to unity.

In an experimental analysis, it is impossible to tell on an event-by-event basis
whether a quark or an antiquark is emitted in the first stage of
eq.~\eqref{chain}, $i.\,e.$ whether the cascade decay was initiated by a
particle $D$ or its antiparticle $\overline{D}$. However, the observable $j\ell$
invariant-mass distribution may depend significantly on the fraction $f$ of
events stemming from $D$ decays versus the fraction $\bar{f}$ of events stemming
from $\overline{D}$ decays, with $f+\bar{f}=1$.

As pointed out in Ref.~\cite{Burns:2008cp}, the ratio of $f$ and $\bar{f}$ is
very difficult to determine without model assumption and thus should be treated
as a free parameter. The $j\ell$ distribution depends on $f$ and $\bar{f}$ only
through the combinations $f|c_L|^2+\bar{f}|c_R|^2 = f \cos^2 \gamma + \bar{f}
\sin^2 \gamma$ and $f|c_R|^2+\bar{f}|c_L|^2 = f \sin^2 \gamma + \bar{f}
\cos^2 \gamma$. It is therefore convenient to introduce the parameter
$\tilde{\gamma}$, defined by \cite{Burns:2008cp}
\begin{align}
\cos^2\tilde{\gamma} &= f \cos^2 \gamma + \bar{f} \sin^2 \gamma, \label{gam1} \\
\sin^2\tilde{\gamma} &= f \sin^2 \gamma + \bar{f} \cos^2 \gamma. \label{gam2}
\end{align}
From the analysis of the $j\ell$ invariant-mass distribution one can only obtain
a constraint on $\tilde{\gamma}$, but not on $\gamma$ and $f$ independently.


\section{Invariant-mass distributions}
\label{dist}

As pointed out above, it is difficult to discriminate experimentally between the
decay chain in Fig.~\ref{fig:diag}, with a quark emitted in the first stage, and its
charge-conjugated version with an antiquark emitted in the first stage, since
both quark and antiquark fragment into jets.
Therefore the only relevant observable invariant-mass distributions are the
$m_{\ell\ell}$ (lepton-lepton) distribution and the $m_{j\ell}$ (jet-lepton)
distribution.

There is no distinction between the two leptons in the three-body decay, in
contrast to the situation when $B$ can be produced on-shell ($i.,e.$ for $m_B <
m_C$) in which case one can define a ``near'' and a ``far'' lepton
\cite{mass,spin,spin2,Miller:2005zp,Burns:2008cp}.

Explicit expressions for the $m_{\ell\ell}$ and $m_{j\ell}$ distributions are
obtained by computing the squared matrix elements for the different spin
configurations $S$=1--11 in Tab.~\ref{tab:spins} and integrating over the
remaining phase space variables. A convenient choice
for the phase space integration is given by
\begin{align}
\frac{1}{\Gamma} \, \frac{d\Gamma}{d m_{\ell\ell}^2} &= N_{\ell\ell}
 \int_{m_{A\ell^-}^{\rm min}}^{m_{A\ell^-}^{\rm max}}
  d m_{A\ell^-}^2 \, |{\cal M}_3|^2,    
  \\
  &\hspace{8em} m_{A\ell^-}^{\rm min,max} = \tfrac{1}{2}
   [m_A^2+m_C^2-m_{\ell\ell}^2\mp\lambda^{1/2}(m_A^2,m_C^2,m_{\ell\ell}^2)],
  \nonumber
\\[1em]
\frac{1}{\Gamma} \, \frac{d\Gamma}{d m_{q\ell^+}^2} &= N_{q\ell}
  \int_{m_A^2}^{m_C^2[1-m_{q\ell^+}^2/(m_D^2-m_C^2)]} dm_{A\ell^-}^2 
  \int_0^{2\pi} d\phi \\
  \nonumber & \hspace{1.5em} \times
  \int_0^{(m_{A\ell^-}^2 - m_A^2)(m_C^2 - m_{A\ell^-}^2)/m_{A\ell^-}^2} 
   dm_{\ell\ell}^2 \,\, \frac{1}{m_C^2-m_{A\ell^-}^2} \, |{\cal M}_4|^2,
\end{align}
where $\lambda(a,b,c) \equiv a^2+b^2+c^2-2(ab+ac+bc)$. In these equations,
${\cal M}_{3,4}$ denote the matrix elements for the 3-body  or (3+1)-body decay
processes, respectively, while $m_{A\ell^-}$ is the invariant mass of particle
$A$ and one of the leptons, and $\phi$ is the angle between the plane spanned
by the lepton-lepton system and the quark in the reference frame of $C$.
The charge of the lepton in $m_{A\ell^-}$ and $m_{q\ell^+}$ has been specified for
definiteness, but one can equally well choose the variables $m_{A\ell^+}$ and
$m_{q\ell^-}$.
$N_{\ell\ell}$ and $N_{q\ell}$ are unspecified normalization constants.

The observable jet-lepton distribution
$d\Gamma/d m_{j\ell}^2$ is obtained from $d\Gamma/d m_{q\ell}^2$
by replacing $\gamma$ with $\tilde{\gamma}$, see eqs.~\eqref{gam1},\eqref{gam2}.

As mentioned above, the endpoints of the invariant-mass distributions can be
used to obtain information about the masses $m_A$, $m_C$ and $m_D$ of the
particles that are produced on-shell in the cascade, while the shapes of the
distributions depend on the couplings and spins of the particles $A$--$D$.
Focusing on the latter, it is convenient to define the distributions 
$d\Gamma/d \hat{m}_{\ell\ell}$ and $d\Gamma/d \hat{m}_{j\ell}$ in terms
of unit-normalized invariant masses
\begin{align}
\hat{m}_{\ell\ell} &\equiv \frac{m_{\ell\ell}}{m_{\ell\ell}^{\rm max}}, &
{m_{\ell\ell}^{\rm max}} &= m_C-m_A, \\
\hat{m}_{j\ell} &\equiv \frac{m_{j\ell}}{m_{j\ell}^{\rm max}}, &
({m_{j\ell}^{\rm max}})^2 &= \frac{1}{m_C^2}(m_D^2-m_C^2)(m_C^2-m_A^2).
\end{align}
For the spin configurations $S$=1--6, the dependence on the coupling parameters
$\alpha,\beta,\tilde{\gamma}$ can be cast into the form
\begin{align}
\frac{1}{\Gamma} \, \frac{d\Gamma}{d \hat{m}_{\ell\ell}} &= \hspace{-1em}
\begin{aligned}[t]
  &(\cos^2\alpha\,\sin^2\beta + \sin^2\alpha\,\cos^2\beta)
  \, f^{(\ell\ell)}_1(\hat{m}_{\ell\ell}^2;\, m_A^2,m_B^2,m_C^2) \\
+ &(\cos^2\alpha\,\cos^2\beta + \sin^2\alpha\,\sin^2\beta)
  \, f^{(\ell\ell)}_2(\hat{m}_{\ell\ell}^2;\, m_A^2,m_B^2,m_C^2) \\ 
+ &(\cos\alpha\,\sin\alpha\,\cos\beta\,\sin\beta)
  \, f^{(\ell\ell)}_3(\hat{m}_{\ell\ell}^2;\, m_A^2,m_B^2,m_C^2),
\end{aligned} \label{all} \displaybreak[0] \\
\frac{1}{\Gamma} \, \frac{d\Gamma}{d \hat{m}_{j\ell}} &= \hspace{-1em}
\begin{aligned}[t]
  &(\cos^2\alpha\,\sin^2\beta\,\cos^2\tilde{\gamma} 
  	+ \sin^2\alpha\,\cos^2\beta\,\sin^2\tilde{\gamma})
  \, f^{(j\ell)}_1(\hat{m}_{j\ell}^2;\, m_A^2,m_B^2,m_C^2,m_D^2) \\
+ &(\cos^2\alpha\,\sin^2\beta\,\sin^2\tilde{\gamma} 
  	+ \sin^2\alpha\,\cos^2\beta\,\cos^2\tilde{\gamma})
  \, f^{(j\ell)}_2(\hat{m}_{j\ell}^2;\, m_A^2,m_B^2,m_C^2,m_D^2) \\
+ &(\cos^2\alpha\,\cos^2\beta\,\cos^2\tilde{\gamma} 
  	+ \sin^2\alpha\,\sin^2\beta\,\sin^2\tilde{\gamma})
  \, f^{(j\ell)}_3(\hat{m}_{j\ell}^2;\, m_A^2,m_B^2,m_C^2,m_D^2) \\
+ &(\cos^2\alpha\,\cos^2\beta\,\sin^2\tilde{\gamma} 
  	+ \sin^2\alpha\,\sin^2\beta\,\cos^2\tilde{\gamma})
  \, f^{(j\ell)}_4(\hat{m}_{j\ell}^2;\, m_A^2,m_B^2,m_C^2,m_D^2) \\
+ &(\cos\alpha\,\sin\alpha\,\cos\beta\,\sin\beta)
  \, f^{(j\ell)}_5(\hat{m}_{j\ell}^2;\, m_A^2,m_B^2,m_C^2,m_D^2), 
\label{aql}
\end{aligned}
\end{align}
where the functions $f^{(\ell\ell)}_i$ and $f^{(j\ell)}_i$ are independent of
the coupling parameters, but they contain the entire kinematical and spin
information, including the dependence on the particle masses.
Note that $f^{(\ell\ell)}_3$ and $f^{(j\ell)}_5$ receive contributions only from the
interference term between the $t$- and $u$-channel diagrams in the upper part of
Fig.~\ref{fig:diag}, see also Ref.~\cite{Csaki:2007xm}.

From eqs.~\eqref{all},\eqref{aql} one can see that without loss of generality
the coupling parameters can be restricted to the intervals $\alpha \in
[-\pi/2,\pi/2],\,\beta,\tilde{\gamma} \in [0,\pi/2]$, as already mentioned in
the previous section.

For $S$=7--11, the $CAZ$ coupling is uniquely fixed up to an overall coupling
constant, so that there is only one term for the lepton-lepton invariant-mass
distribution. However, there are two possible terms for the jet-lepton  invariant-mass
distribution:
\begin{align}
\frac{1}{\Gamma} \, \frac{d\Gamma}{d \hat{m}_{\ell\ell}} &= 
  f^{(\ell\ell)}(\hat{m}_{\ell\ell}^2;\, m_A^2,m_Z^2,m_C^2),
\label{bll}
\\
\frac{1}{\Gamma} \, \frac{d\Gamma}{d \hat{m}_{j\ell}} &= 
  f^{(j\ell)}_{\rm S}(\hat{m}_{j\ell}^2;\, m_A^2,m_Z^2,m_C^2,m_D^2) 
+ \cos 2\tilde{\gamma}
  \, f^{(j\ell)}_{\rm A}(\hat{m}_{j\ell}^2;\, m_A^2,m_Z^2,m_C^2,m_D^2),
\label{bql}  
\end{align}
The lepton-lepton distribution $d\Gamma/d \hat{m}_{\ell\ell}$ can be expressed in
terms of compact analytical formulae. On the other hand, the analytical results
for $d\Gamma/d\hat{m}_{q\ell}$ are very lengthy, so that instead we chose to perform
the last integration step (over $m_{A\ell^-}^2$) numerically. 

Explicit expressions for the functions $f^{(xy)}_i$ are available for free 
download (see appendix).
Figs.~\ref{fig:ll}--\ref{fig:qllz} depict the distribution functions for a
sample mass spectrum. In the figures, the overall normalization
constants have been fixed by requiring that $f^{(\ell\ell)}_1$, $f^{(j\ell)}_1$,
$f^{(\ell\ell)}$, and $f^{(j\ell)}_{\rm S}$ are unit-normalized. The right
column of Fig.~\ref{fig:ll} also illustrates how the distributions vary with the
mass $m_B$ of the off-shell intermediate particle $B$, for the example of the
spin configuration $S$=1.

\begin{figure}[p]
\begin{tabular}{@{}r@{\hspace{2mm}}r@{}}
\epsfig{figure=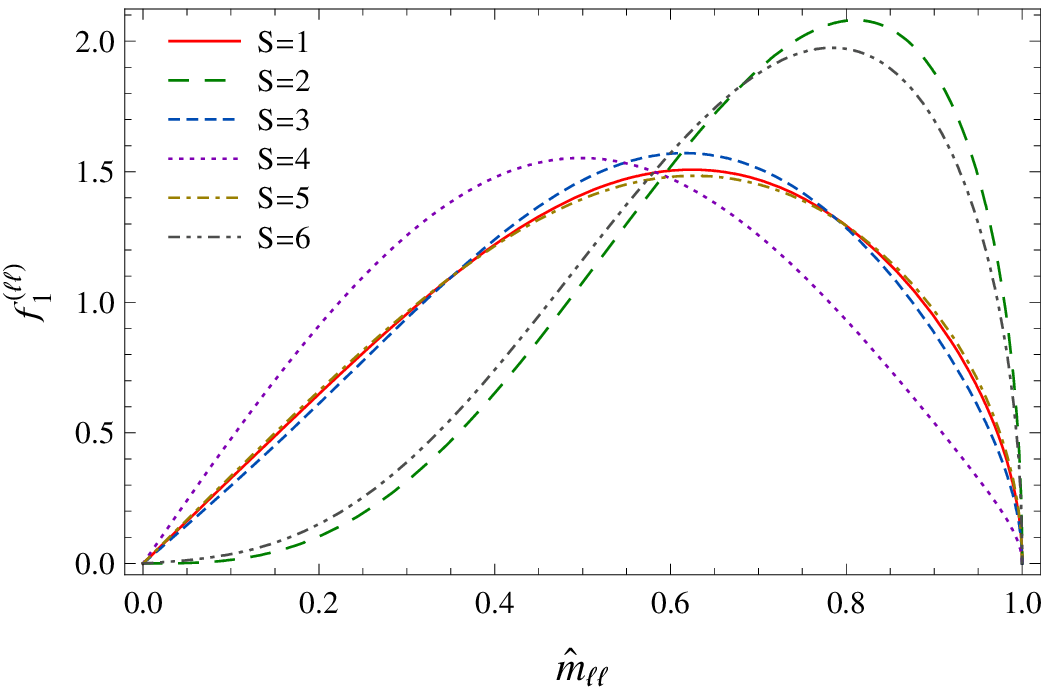, height=5.3cm} &
\epsfig{figure=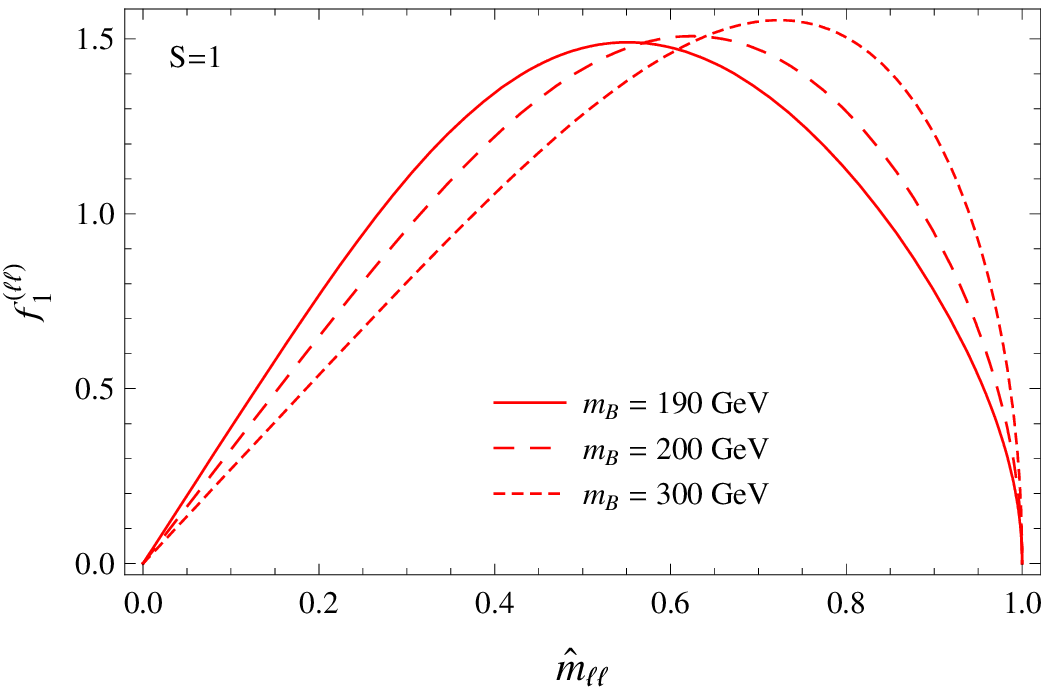, height=5.3cm} \\[1ex]
\raisebox{.5mm}{\epsfig{figure=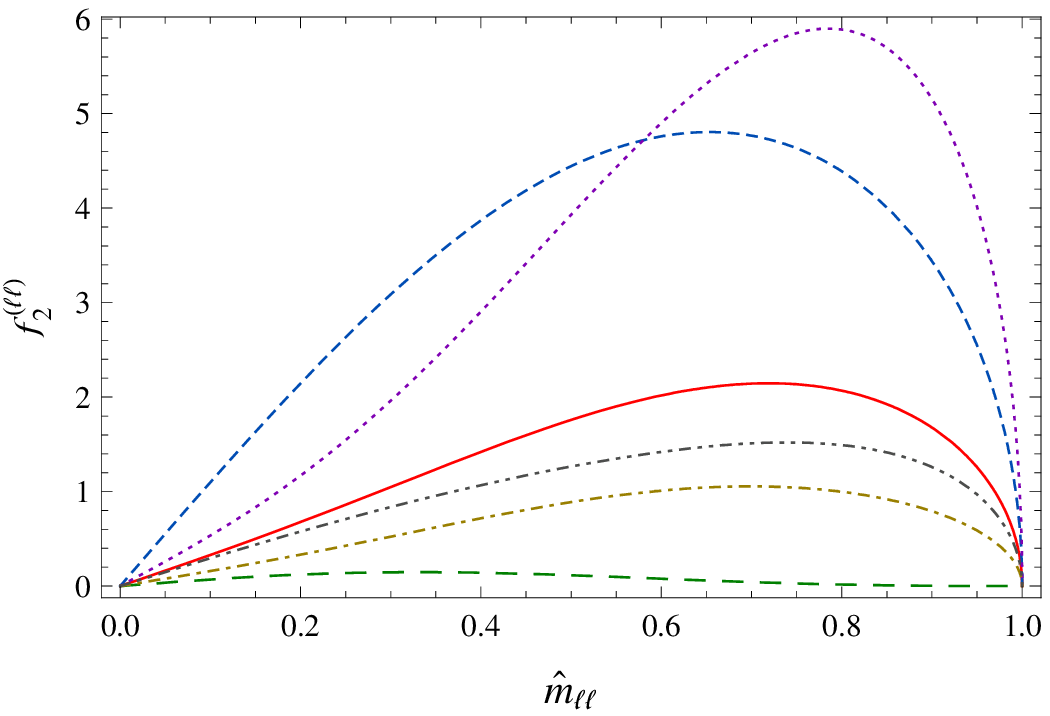, height=5.3cm}} &
\epsfig{figure=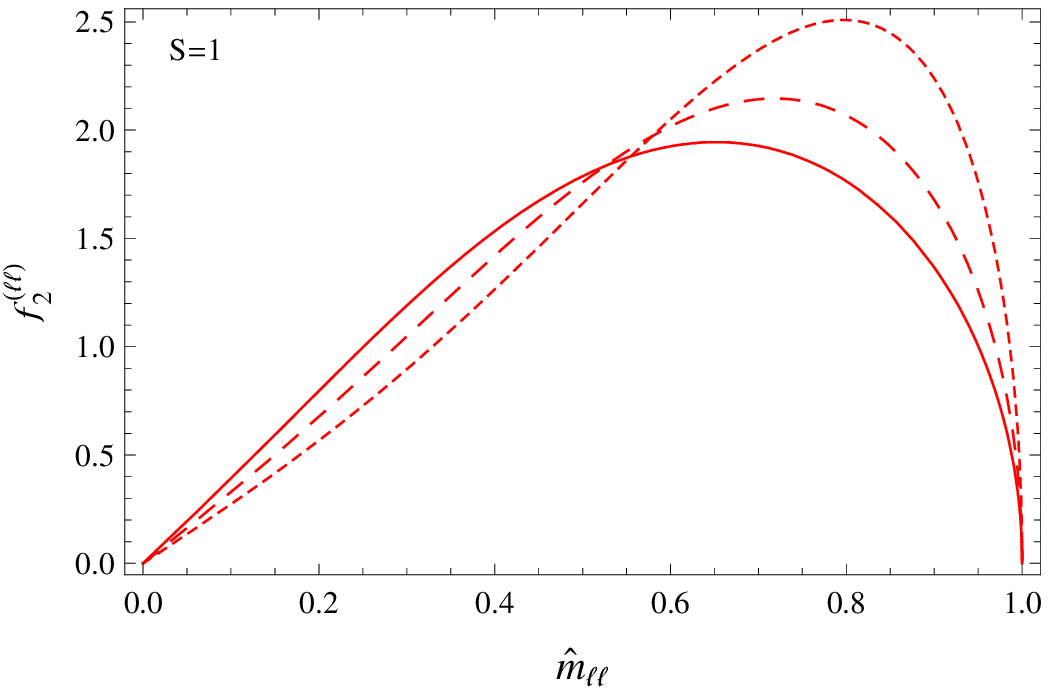, height=5.3cm} \\[1ex]
\epsfig{figure=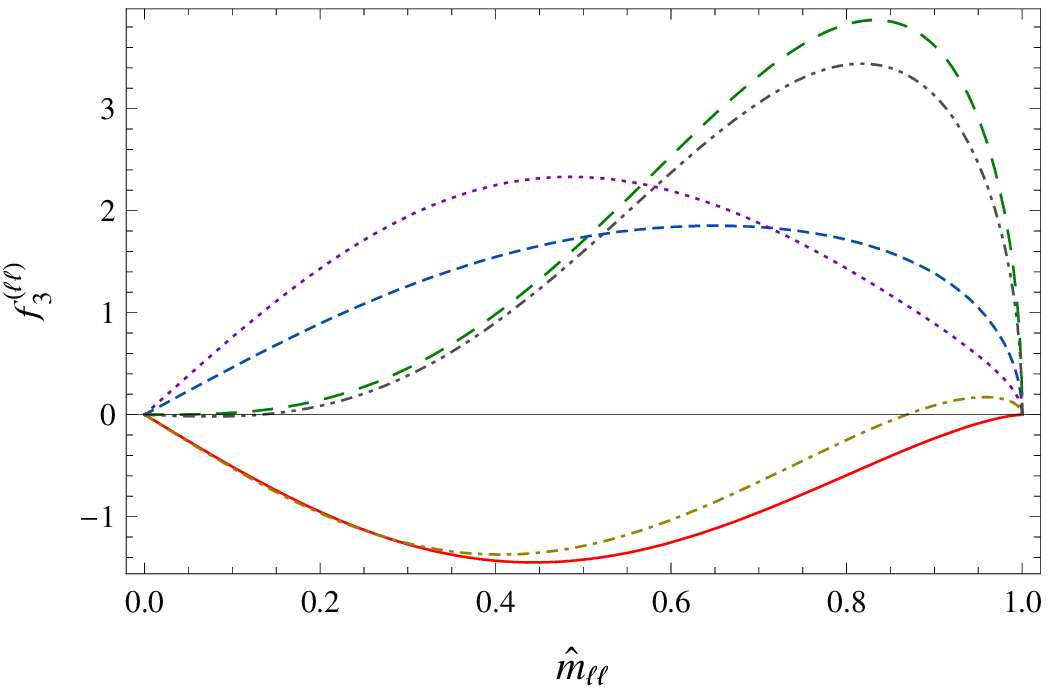, height=5.3cm} &
\epsfig{figure=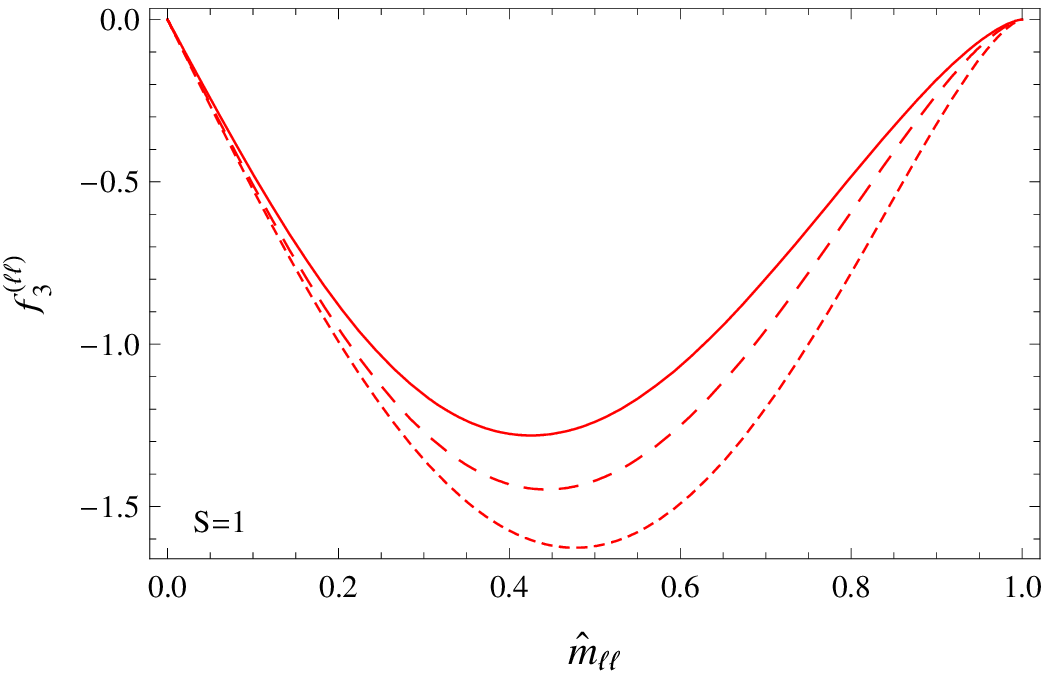, height=5.3cm}
\end{tabular}
\vspace{-4mm}
\mycaption{Left: Distribution functions $f^{(\ell\ell)}_i$ ($i=\,$1,...,3) for the spin
configurations $S$=1--6, for $m_B =\,$200~GeV. 
Right: Dependence of $f^{(\ell\ell)}_i$ ($i=\,$1,...,3) on the mass $m_B$ of the intermediate
particle for the case $S$=1. The other mass parameters have been chosen as
$m_C =\,$184~GeV and $m_A =\,$98~GeV. In these plots the overall 
normalization has been fixed by normalizing $f^{(\ell\ell)}_1$ to unity.}
\label{fig:ll}
\end{figure}
\begin{figure}[p]
\begin{tabular}{@{}r@{\hspace{2mm}}r@{}}
\epsfig{figure=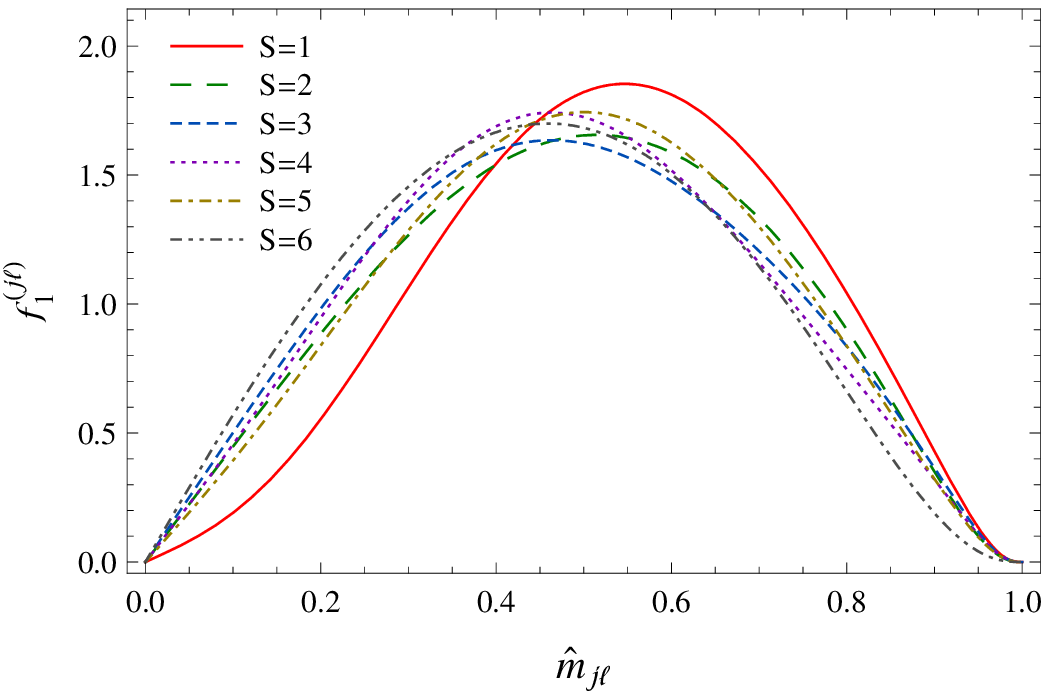, height=5.3cm} &
\epsfig{figure=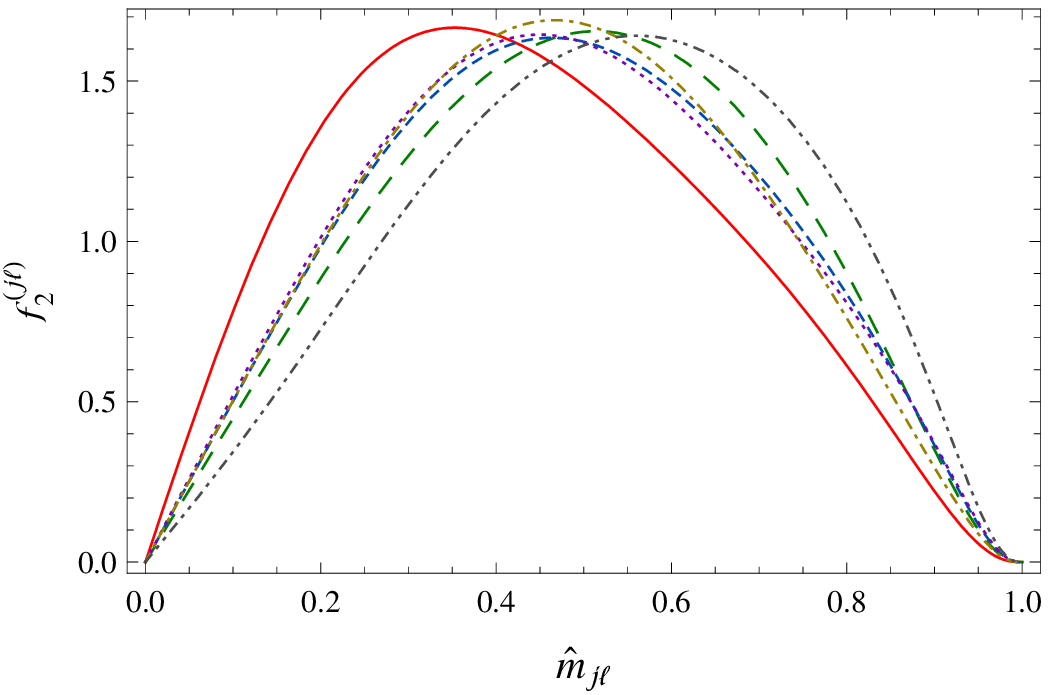, height=5.3cm} \\[1ex]
\epsfig{figure=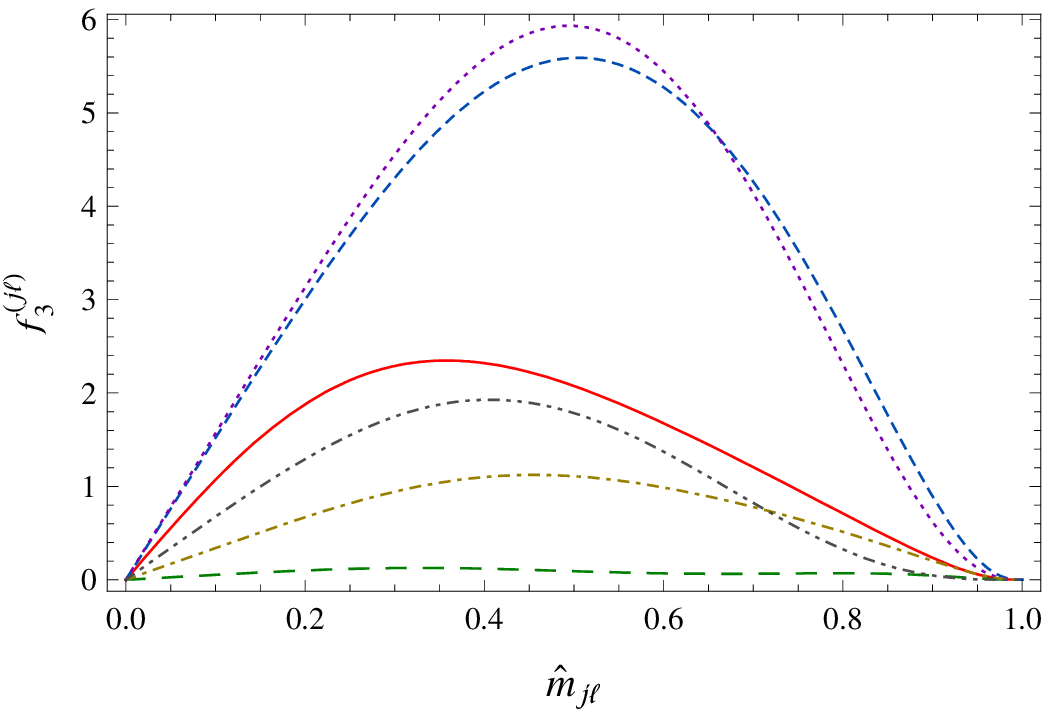, height=5.3cm} &
\epsfig{figure=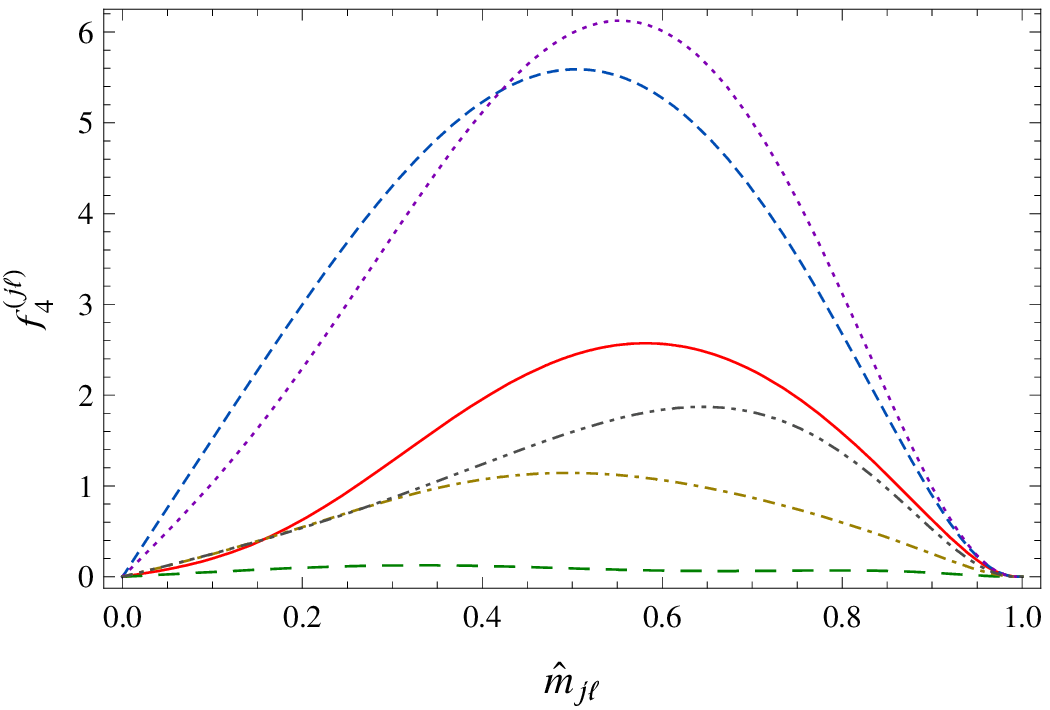, height=5.3cm} \\[1ex]
\epsfig{figure=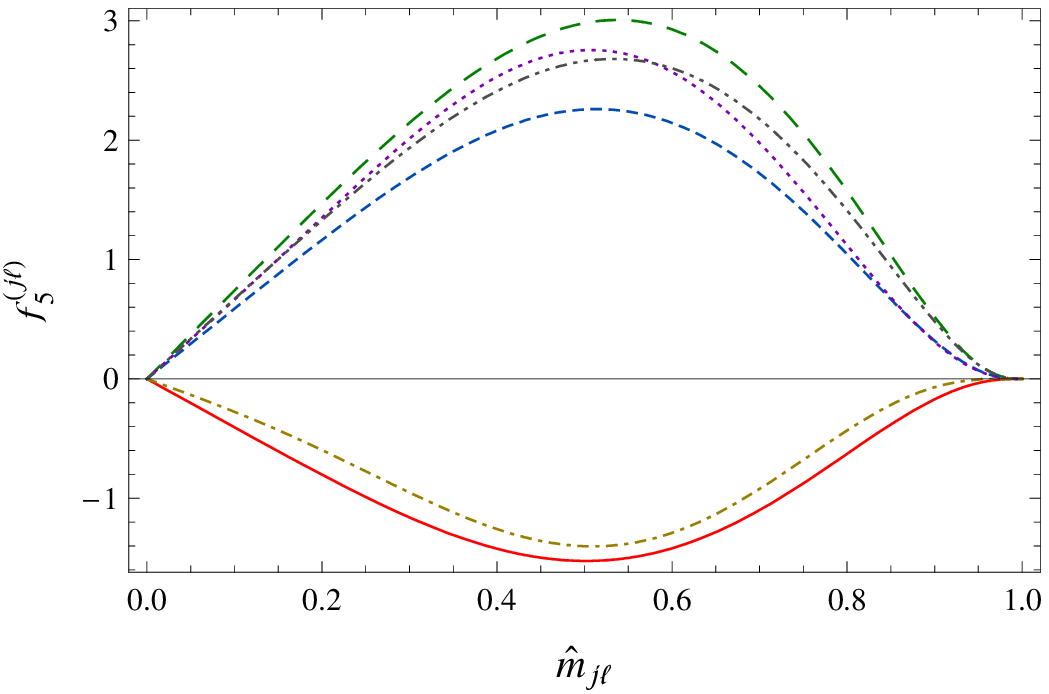, height=5.3cm} &
\end{tabular}
\vspace{-4mm}
\mycaption{Distribution functions $f^{(j\ell)}_i$ ($i=\,$1,...,5) for the spin
configurations $S$=1--6.  The mass parameters have been chosen as
$m_D=\,$565~GeV, $m_C =\,$184~GeV, $m_B =\,$200~GeV and $m_A =\,$98~GeV. In
these plots the overall  normalization has been fixed by normalizing
$f^{(j\ell)}_1$ to unity.}
\label{fig:ql}
\end{figure}
\begin{figure}[t]
\centering
\epsfig{figure=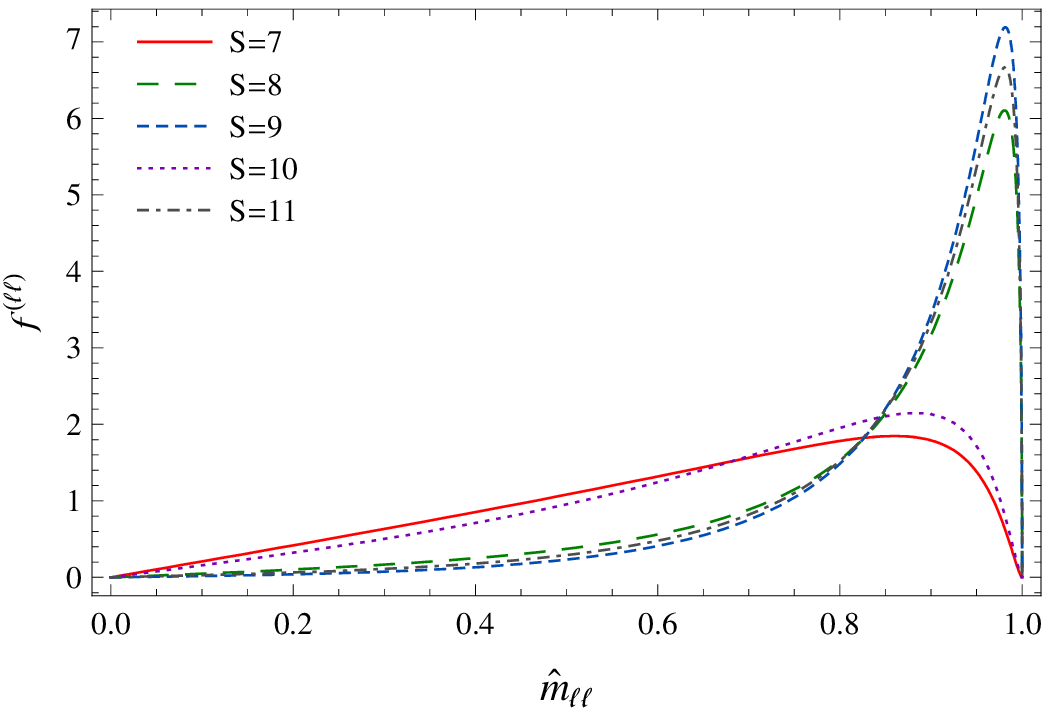, height=5.3cm} \\[1ex]
\epsfig{figure=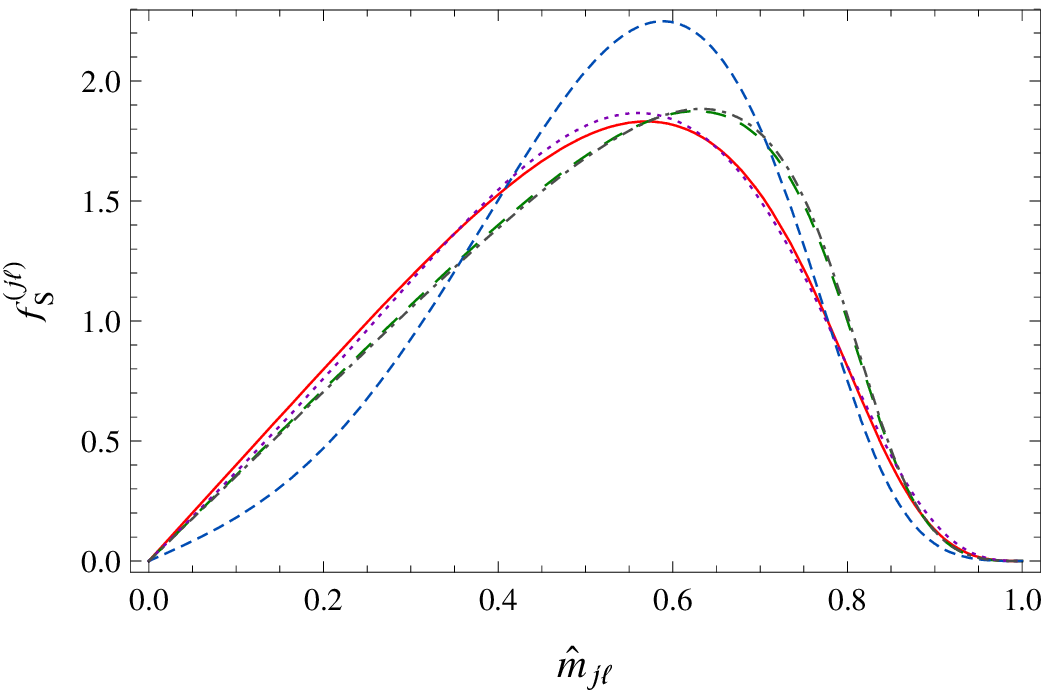, height=5.3cm} \hfill
\epsfig{figure=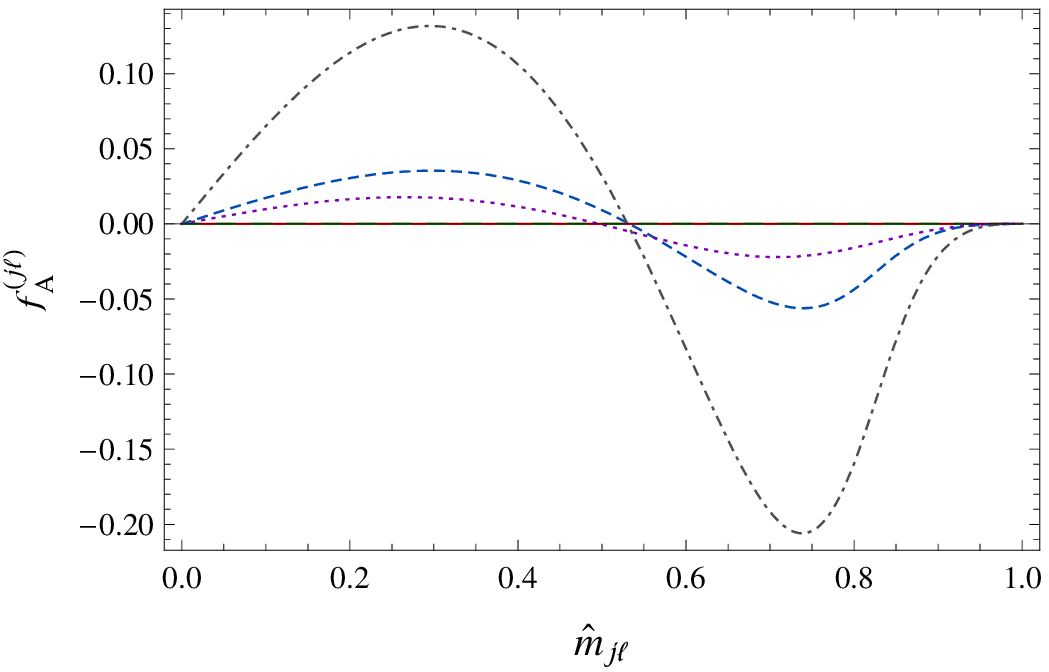, height=5.3cm} 
\vspace{-8mm}
\mycaption{Distribution functions $f^{(\ell\ell)}$ and 
$f^{(j\ell)}_{\rm S,A}$ for the spin configurations $S$=7--11.  The mass
parameters have been chosen as $m_D=\,$565~GeV, $m_C =\,$184~GeV  and $m_A
=\,$98~GeV. In these plots the overall normalization has been fixed by
normalizing $f^{(\ell\ell)}$ and $f^{(j\ell)}_{\rm S}$ to unity.}
\label{fig:qllz}
\end{figure}


\section{Analysis method}
\label{ana}

In this section we will discuss the determination of the spins and couplings
parameters of the new particles, as well as the mass of the off-shell particle
$B$, by fitting the theoretically calculated distributions to experimental data.
The general procedure will be outlined in the next subsection, while its
application will be demonstrated in subsection \ref{num} for two concrete
numerical examples.

\subsection{Conceptual procedure}
\label{proc}

The analysis is based on a binned $\chi^2$ fit for the $\ell\ell$ and $j\ell$
distributions. In this fit, the binned histogram for the data is compared with
theoretical histograms obtained by numerically integrating the functions
$f^{(\ell\ell)}_i$ and $f^{(j\ell)}_i$, defined in the previous section, over
the interval covered by each bin.
In the fit, the coupling parameters $\alpha,\,\beta,\,\tilde{\gamma}$ and the
mass $m_B$ are kept as free parameters. Varying over these parameters and the
spin configuration $S$, the best-fit result is found as the set of numbers
$\{S,\alpha,\beta,\tilde{\gamma},m_B\}$ that minimizes the $\chi^2$ value.

During the fit procedure, for every given choice of the parameters
$\{S,\alpha,\beta,\tilde{\gamma},m_B\}$, the theoretical histograms for the
$\ell\ell$ and $j\ell$ distributions are normalized such that the total number
of events in the theoretical histogram agrees with the number of events in the
data histogram. In practice, this normalization is most easily carried out
numerically.

In general, it may happen that there is not a unique solution for the minimum
$\chi^2$ value, but instead several degenerate best-fit points are obtained. In
such a situation, the coupling parameters $\alpha,\,\beta,\,\tilde{\gamma}$
and/or the spin assignment $S$ cannot be determined uniquely from the observable
distributions of the decays \eqref{3b},\eqref{chain} alone.

\subsection{Numerical examples}
\label{num}

To illustrate the fitting procedure, its application is demonstrated by
performing a fit to mock-up data histograms.
This section is based on the parton-level description of the decay
processes \eqref{3b},\eqref{chain} as described in the previous sections, thus
neglecting issues such as backgrounds, jet combinatorics and energy smearing,
which are relevant in a realistic experimental setup. However, earlier studies
\cite{mass,lhclc} have shown that, for mass parameters similar to the ones chosen here, it is
possible to obtain a clean, almost background-free sample of signal events with
relatively simple selection cuts.

Let us consider two sample choices for the hypothetical data:
\begin{list}{}{\setlength{\leftmargin}{6.5em}\setlength{\rightmargin}{1em}
               \setlength{\labelsep}{\bigskipamount}
	       \setlength{\labelwidth}{5.5em}}
\item[\bf ``Data'' A:]
$S=1,\;\;\; \alpha=0,\;\beta=\pi/2,\;\tilde{\gamma}=0,$\\
$m_D=565$~GeV, $m_C=184$~GeV, $m_B=200$~GeV, $m_A=98$~GeV\\[.5ex] 
(corresponding to the MSSM decay chain $\tilde{q}_L \to \tilde{\chi}^0_2 \to
\tilde{l}^*_L \to \tilde{\chi}^0_1$);
\item[\bf ``Data'' B:]
$S=11,\;\;\; \tilde{\gamma}=0,$\\
$m_D=565$~GeV, $m_C=184$~GeV, $m_A=98$~GeV\\[.5ex] 
(corresponding to the MSSM decay chain $\tilde{q}_L \to \tilde{\chi}^0_2 \to
\tilde{\chi}^0_1$).
\end{list}
For each case, we have computed ``data'' histograms with 10 bins each for the
$\hat{m}_{\ell\ell}$ and the $\hat{m}_{j\ell}$ distributions, corresponding to a
total of 1000 events. Then we have performed a $\chi^2$ fit of the theoretical
distribution functions to these fake ``data'' histogram for each of the spin
configurations $S$=1--11, searching for the minimum $\chi^2$ value as a function
of the parameters $\alpha,\,\beta,\,\tilde{\gamma}$, and $m_B$\footnote{For the
spin configurations $S$=7--11, the non-zero $Z$-boson width has been included
although its numerical impact is not very important for the masses chosen
here.}.

\begin{table}[t]
a) ``Data'' A, using only $\hat{m}_{\ell\ell}$ distribution:
\vspace{-1.5em}
\begin{center}
\renewcommand{\arraystretch}{1.1}
\begin{tabular}[t]{|l|c|ccc|}
\hline
&& \multicolumn{3}{c|}{best-fit parameters} \\[-.5ex]
$S$ & $\min\chi^2$ & $\alpha$ & $\beta$ & $m_B$ [GeV] \\
\hline
\hline
1 [SFSF] & 0.00 & \phantom{$-$}0.00 & 1.57 & 200.0 \\
\hline
2 [FSFS] & 0.00 & $-$1.22 & 1.05 & 209\phantom{.0} \\
\hline
3 [FSFV] & 0.00 & $+$1.14 & 0.43 & 197.7 \\
\hline
4 [FVFS] & 0.27 & $-$1.34 & 0.23 & 216\phantom{.0} \\
\hline
5 [FVFV] & 0.05 & $-$0.38 & 0.38 & 197\phantom{.0} \\
\hline
6 [SFVF] & 0.05 & $-$0.65 & 0.92 & 191.3 \\
\hline
\end{tabular}
\hspace{\smallskipamount}
\begin{tabular}[t]{|l|r|}
\hline
&\\[-.5ex]
\phantom{1}$S$ & $\min\chi^2$ \\
\hline
\hline
\phantom{1}7 [FSS] & 140 \\
\hline
\phantom{1}8 [FSV] & 3100 \\
\hline
\phantom{1}9 [FVS] & 4200 \\
\hline
10 [FVV] & 290 \\
\hline
11 [SFF] & 3700 \\
\hline
\end{tabular}
\end{center}
b) ``Data'' A, using both $\hat{m}_{\ell\ell}$ and $\hat{m}_{j\ell}$
distributions:
\vspace{-1.5em}
\begin{center}
\renewcommand{\arraystretch}{1.2}
\begin{tabular}[t]{|l|r|cccc|}
\hline
&& \multicolumn{4}{c|}{best-fit parameters} \\[-.5ex]
$S$ & $\min\chi^2$ & $\alpha$ & $\beta$ & $\tilde{\gamma}$ & $m_B$ [GeV] \\
\hline
\hline
1 [SFSF] & 0 & \phantom{$-$}0.00 & 1.57 & 0.00 & 200.0 \\
\hline
2 [FSFS] & 150 & $-$0.08 & 0.07 & 1.57 & 754\phantom{.0} \\
\hline
3 [FSFV] & 87 & $\pm$1.57 & 1.57 & 0.29 & 210\phantom{.0} \\
\hline
4 [FVFS] & 48 & $\pm$1.19 & 0.00 & 1.57 & 220\phantom{.0} \\
\hline
5 [FVFV] & 46 & $-$0.93 & 0.25 & 1.57 & 224\phantom{.0} \\
\hline
6 [SFVF] & 37 & $-$0.50 & 0.53 & 1.57 & 197.4 \\
\hline
\end{tabular}
\hspace{\smallskipamount}
\begin{tabular}[t]{|l|r|c|}
\hline
&& best-fit \\[-.5ex]
\phantom{1}$S$ & $\min\chi^2$ & $\tilde{\gamma}$ \\
\hline
\hline
\phantom{1}7 [FSS] & 200 & ? \\
\hline
\phantom{1}8 [FSV] & 3100 & ? \\
\hline
\phantom{1}9 [FVS] & 4300 & 0.39 \\
\hline
10 [FVV] & 330 & 1.57 \\
\hline
11 [SFF] & 3700 & 1.08 \\
\hline
\end{tabular}
\end{center}
\vspace{-1ex}
\mycaption{Results for fitting all spin configurations $S$=1--11 to (a) the
$\hat{m}_{\ell\ell}$ distribution only, and (b) the $\hat{m}_{\ell\ell}$ and
$\hat{m}_{j\ell}$ distributions together, using scenario ``data'' A for the
mock-up data histograms. Shown are the minimum $\chi^2$ (rounded to two
significant digits) for each spin
configuration, as well as the parameter values for which this minimal value is
attained. ``?'' indicates that the $\chi^2$ value is independent of that parameter. 
The numbers correspond to 1000 events.}
\label{tab:ll1}
\end{table}

\begin{table}[t]
a) ``Data'' B, using only $\hat{m}_{\ell\ell}$ distribution:
\vspace{-1.5em}
\begin{center}
\renewcommand{\arraystretch}{1.1}
\begin{tabular}[t]{|l|r|ccc|}
\hline
&& \multicolumn{3}{c|}{best-fit parameters} \\[-.5ex]
$S$ & $\min\chi^2$ & $\alpha$ & $\beta$ & $m_B$ [GeV] \\
\hline
\hline
1 [SFSF] & 1200 & $+$0.79 & 0.79 & $\infty$ \\
\hline
2 [FSFS] &  670 & ? & ? & $\infty$ \\
\hline
3 [FSFV] & 2200 & ? & ? & $\infty$ \\
\hline
4 [FVFS] & 1100 & ? & ? & $\infty$ \\
\hline
5 [FVFV] &  720 & \multicolumn{2}{c}{$\alpha=\beta=\,?$} & $\infty$ \\
\hline
6 [SFVF] &  740 & $\pm$1.57 & 0.00 & $\infty$ \\
\hline
\end{tabular}
\hspace{\smallskipamount}
\begin{tabular}[t]{|l|r|}
\hline
&\\[-.5ex]
\phantom{1}$S$ & $\min\chi^2$ \\
\hline
\hline
\phantom{1}7 [FSS] & 1600\phantom{.00} \\
\hline
\phantom{1}8 [FSV] & 16\phantom{.00} \\
\hline
\phantom{1}9 [FVS] & 8.7\phantom{0} \\
\hline
10 [FVV] & 1100\phantom{.00} \\
\hline
11 [SFF] & 0.00 \\
\hline
\end{tabular}
\end{center}
b) ``Data'' B, using both $\hat{m}_{\ell\ell}$ and $\hat{m}_{j\ell}$
distributions:
\vspace{-1.5em}
\begin{center}
\renewcommand{\arraystretch}{1.2}
\begin{tabular}[t]{|l|r|cccc|}
\hline
&& \multicolumn{4}{c|}{best-fit parameters} \\[-.5ex]
$S$ & $\min\chi^2$ & $\alpha$ & $\beta$ & $\tilde{\gamma}$ & $m_B$ [GeV] \\
\hline
\hline
1 [SFSF] & 1200 & $+$0.78 & 0.77 & 0.00 & $\infty$ \\
\hline
2 [FSFS] &  690 & ? & ? & ? & $\infty$ \\
\hline
3 [FSFV] & 2300 & ? & ? & ? & $\infty$ \\
\hline
4 [FVFS] & 1100 & $\pm$1.25 & 0.43 & 1.32 & $\infty$ \\
\hline
5 [FVFV] &  750 & $+$0.46 & 0.46 & 1.57 & $\infty$ \\
\hline
6 [SFVF] &  760 & $\pm$1.57 & 0.00 & ? & $\infty$ \\
\hline
\end{tabular}
\hspace{\smallskipamount}
\begin{tabular}[t]{|l|r|c|}
\hline
&& best-fit \\[-.5ex]
\phantom{1}$S$ & $\min\chi^2$ & $\tilde{\gamma}$ \\
\hline
\hline
\phantom{1}7 [FSS] & 1600\phantom{.00} & ? \\
\hline
\phantom{1}8 [FSV] & 25\phantom{.00} & ? \\
\hline
\phantom{1}9 [FVS] & 59\phantom{.00} & 0.00 \\
\hline
10 [FVV] & 1100\phantom{.00} & 0.00 \\
\hline
11 [SFF] & 0.00 & 0.00 \\
\hline
\end{tabular}
\end{center}
\vspace{-1ex}
\mycaption{Same as Fig.~\ref{tab:ll1}, but using ``data'' B for the
mock-up data histograms.}
\label{tab:ll2}
\end{table}

The results are shown in Tables~\ref{tab:ll1} and~\ref{tab:ll2}. From
Tab.~\ref{tab:ll1} one can see that when only information about the
$\hat{m}_{\ell\ell}$ distribution is available, it is difficult to distinguish
the ``data'' A (based on the spin configuration $S$=1) from the spin
configurations $S$=2--6. The underlying reason is that for each of these spin
configurations there are three unknown continuous parameters, $\alpha$, $\beta$
and $m_B$, which can be adjusted so as to mimic the data distribution.

On the other hand, the spin configurations $S$=7--11 can be distinguished from
``data'' A with high significance, using only the $\hat{m}_{\ell\ell}$
distribution. This is a consequence of the fact that there are no free
parameters to adjust in $d\Gamma/d\hat{m}_{\ell\ell}$ for $S$=7--11, and that
these spin configurations correspond to a different diagram topology (Topology
II in Fig.~\ref{fig:diag} instead of topology I).

If both the $\hat{m}_{\ell\ell}$ and $\hat{m}_{j\ell}$ distributions are
included in the fit, all possible spin configurations can be discriminated with
at least six standard deviations, for the given number of 1000 events. 

For the second example, it is evident from Tab.~\ref{tab:ll2} that ``data'' B
can be distinguished from all other spin configurations $S$=1--10 by just using
the $\hat{m}_{\ell\ell}$ distribution. In fact, for all combinations except
$S$=8 and $S$=9 the significance for this discrimination is very high and is not
improved substantially by including the $\hat{m}_{j\ell}$ distribution in the
fit. Also note that the best-fit results for $S$=1--6 are obtained for very
large values of $m_B$, since increasing values of $m_B$ shift the
$\hat{m}_{\ell\ell}$ distribution toward larger values of $\hat{m}_{\ell\ell}$,
see Fig.~\ref{fig:ll}~(right), leading to better agreement with the reference
case $S$=11, see Fig~\ref{fig:qllz}.

\begin{figure}[t]
\epsfig{figure=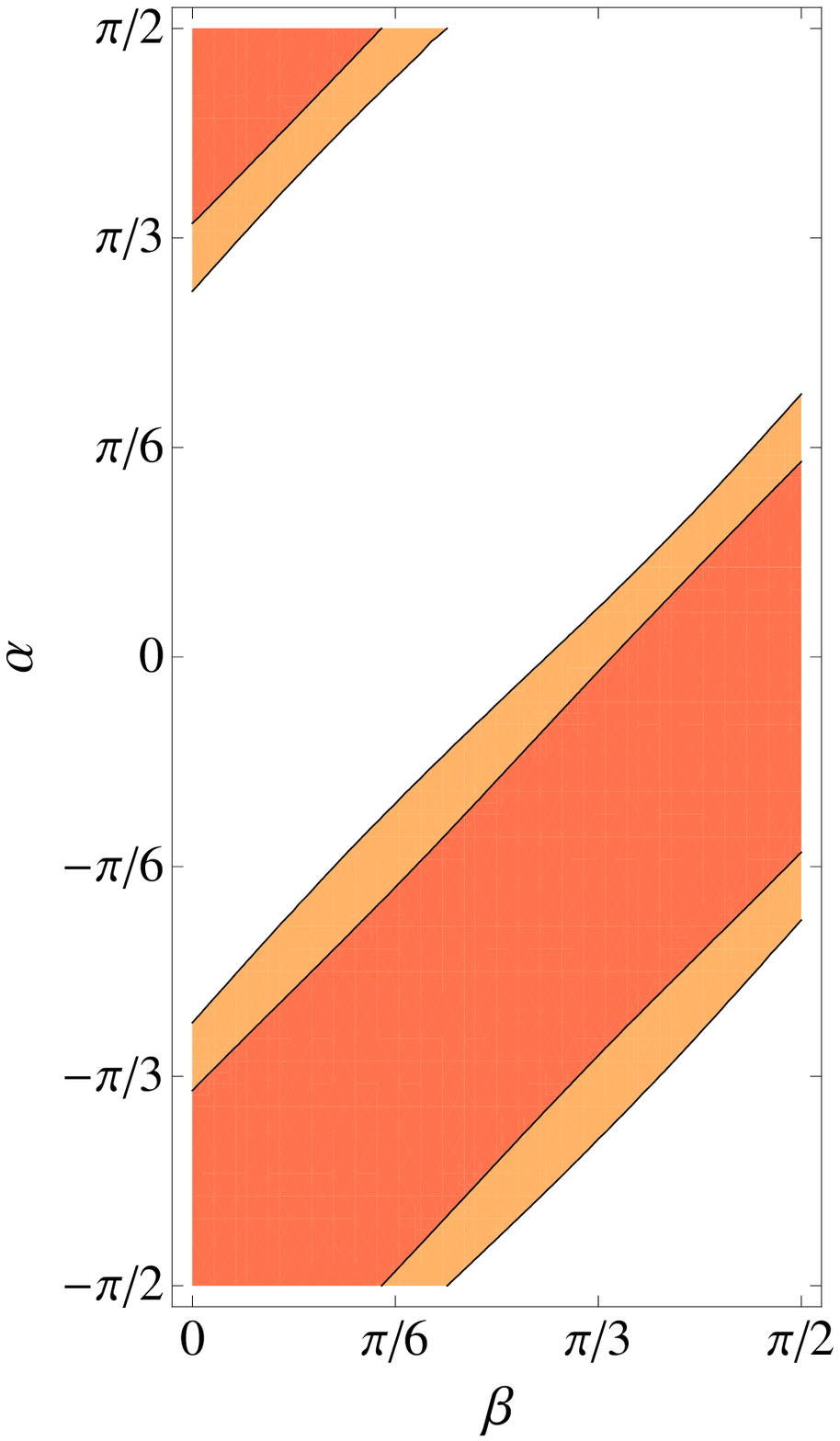, height=7cm, bb=0 5 300 511}%
\rput(-0.6,6.5){(a)}\hspace{1mm}%
\epsfig{figure=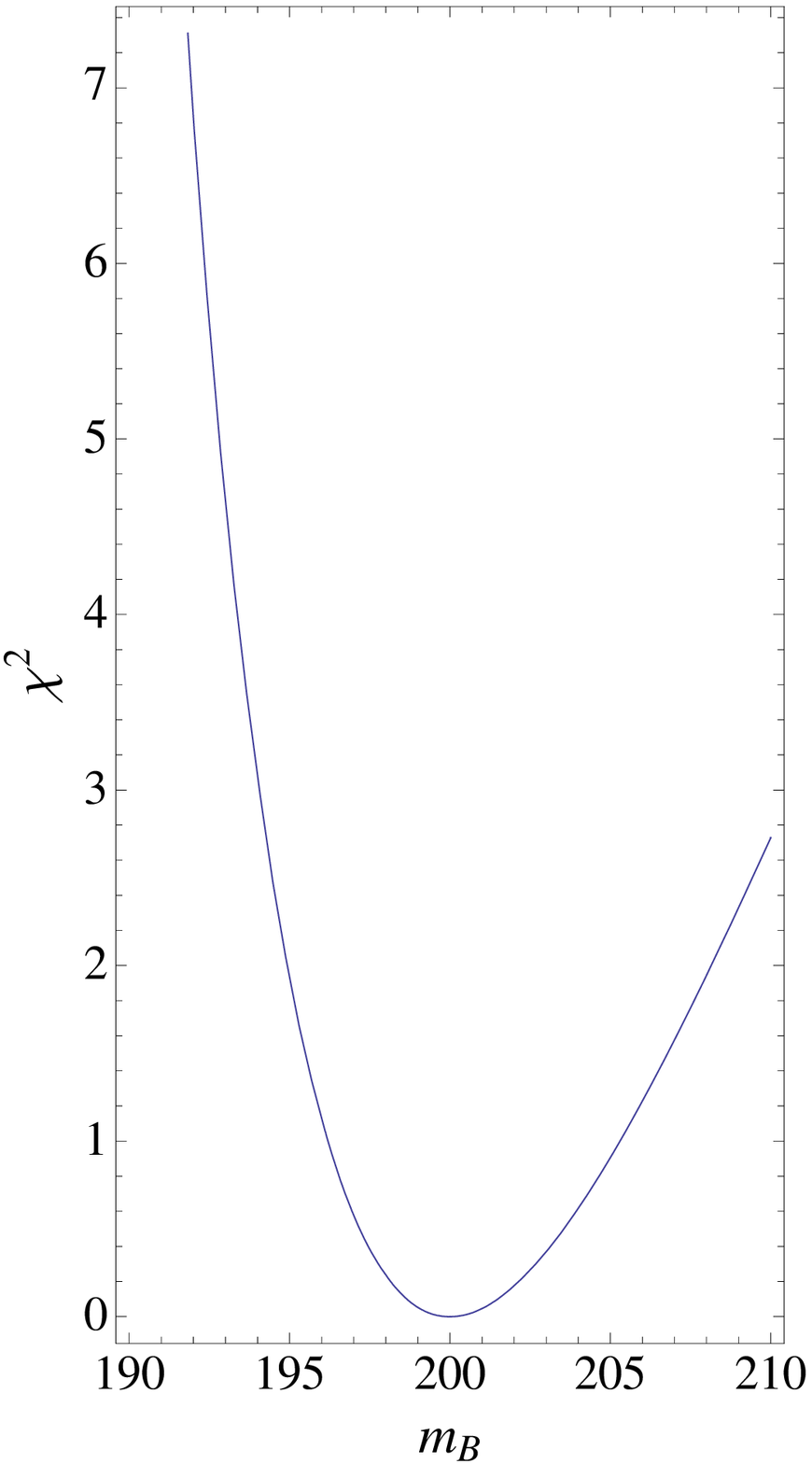, height=7cm}%
\rput(-0.6,6.5){(b)}
\hfill
\epsfig{figure=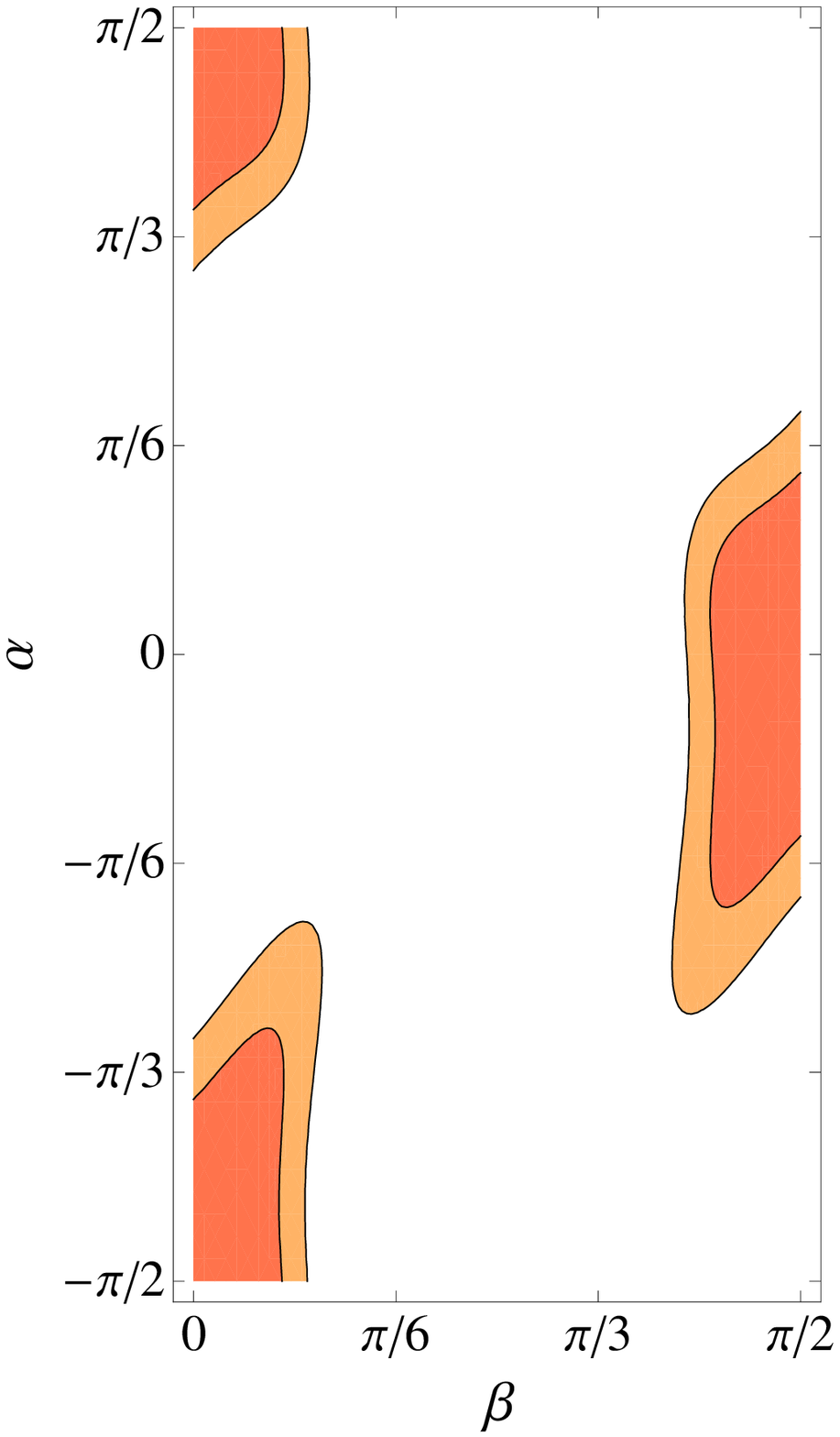, height=7cm, bb=0 5 300 511}%
\rput(-0.6,6.5){(c)}\hspace{1mm}%
\epsfig{figure=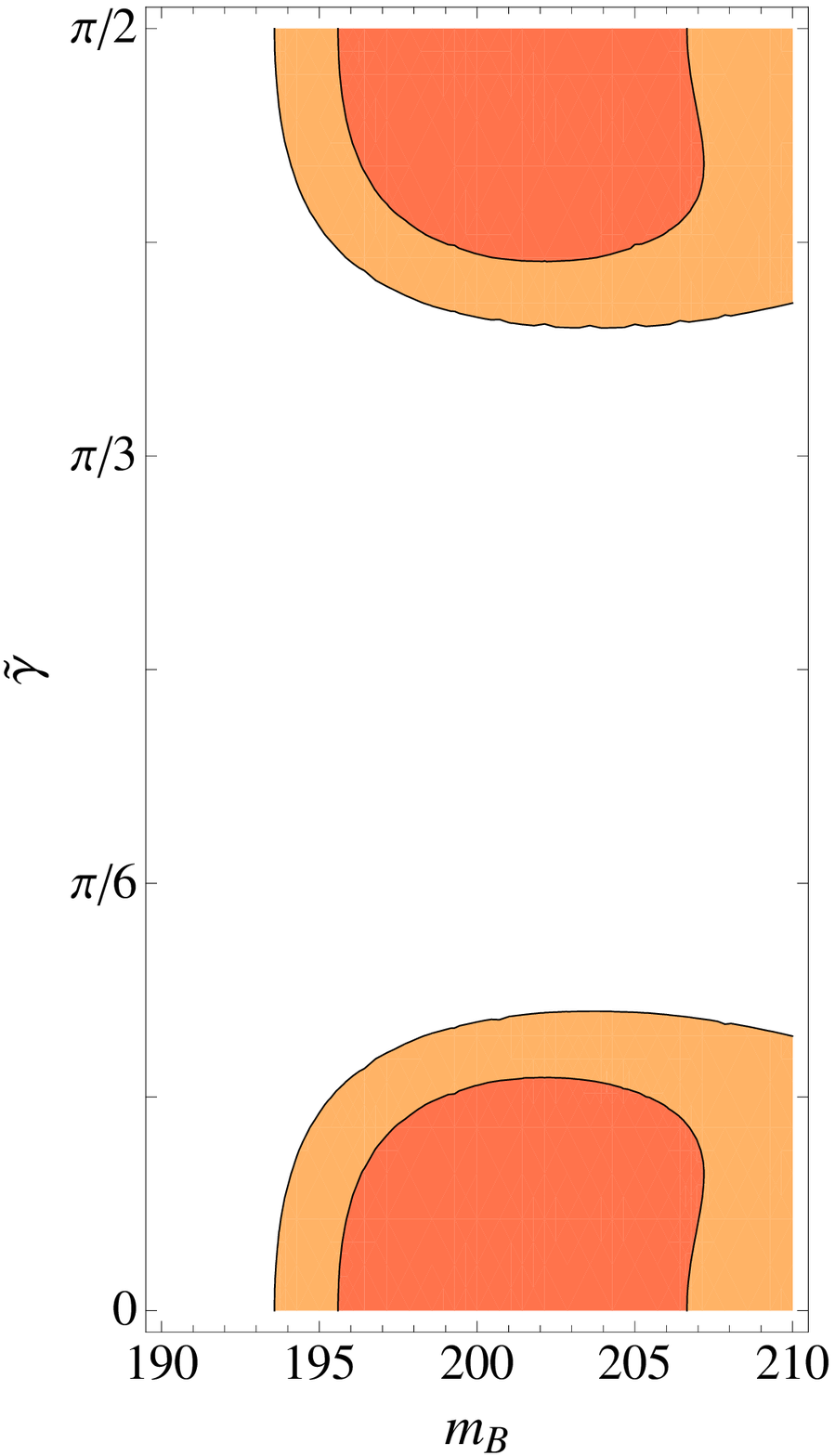, height=7cm}%
\rput(-3,6.5){(d)}
\mycaption{Determination of the parameters $\alpha,\,\beta,\,\tilde{\gamma}$,
and $m_B$ using only the
$\hat{m}_{\ell\ell}$ distribution (a,b), and using both the $\hat{m}_{\ell\ell}$ and
$\hat{m}_{j\ell}$ distributions (c,d). The dark/light bands in the figures
correspond to the 68\%/95\% confidence-level regions. The plots correspond to a
sample of 1000 events for the scenario ``Data'' A.}
\label{fig:parafit}
\end{figure}

\vspace{\medskipamount}
In addition to the spin determination, the couplings of the new particles and
the mass of the off-shell $B$ particle can in principle be extracted from the
fit to the invariant-mass distributions. This is shown in Fig.~\ref{fig:parafit}
for the example of ``data'' A. The panels (a) and (b) in the figure depict the
constraints on $\alpha$, $\beta$ and $m_B$ obtained from fitting the
$\hat{m}_{\ell\ell}$ distribution alone, assuming that $S$=1 is the correct spin
configuration. If a fit to both the $\hat{m}_{\ell\ell}$ and $\hat{m}_{j\ell}$
distributions is performed, one obtains the results in panels (c) and (d). As
evident from the plots, the inclusion of the $\hat{m}_{j\ell}$
distribution does not only lead to a constraint on $\tilde{\gamma}$ (which
cannot be obtained from $d\Gamma/d\hat{m}_{\ell\ell}$), but also to improved
bounds on $\alpha$ and $\beta$.

However, the fit results for the coupling parameters always have a two-fold
degeneracy, since the invariant-mass distributions,
eqs.~\eqref{all}--\eqref{bql}, are invariant under the transformation
$\{\alpha,\,\beta,\,\gamma\} \to \{\text{sign}\,\alpha\,(\frac{\pi}{2}-|\alpha|),\,\frac{\pi}{2}-\beta,
\,\frac{\pi}{2}-\gamma\}$. 

\vspace{2em}


\section{Summary}
\label{summ}

In this paper, a general analysis of three-body decays of the form $C \to \ell^+
\ell^- A$, leading to a pair of opposite-sign leptons and one invisible particle
$A$, has been presented.  This decay process can occur in many proposed
new-physics models, either from direct production of the particle $C$ at the
LHC, or from a  cascade decay of the type $D \to 
\stackrel{\text{\tiny (--)}}{q}\! C \to
\stackrel{\text{\tiny (--)}}{q}\! \ell^+ \ell^- A$, both of which
have been studied here.

No assumptions about the masses, spins and couplings of the participating
new-physics particles have been made, including the off-shell particle $B$ mediating
the three-body decay. Instead, all possible spin configurations and coupling
form factors have been considered. Experimentally, the masses, spins and
coupling parameters may be determined from measuring the invariant-mass
distributions of the visible decay products.

In the present case, there are two independent distributions, one with respect
to the di-lepton ($\ell^+\ell^-$) invariant mass, and the other with respect to
the jet-lepton ($j\ell^\pm$) invariant mass. Results for both have been obtained
in terms of relatively compact analytical functions or one-dimensional integral
representations.

In two concrete numerical examples, it has been tested how well the properties
of the new-physics particles $A$, $B$, $C$ and $D$ can be determined from these
two invariant-mass distributions. It turns out that the di-lepton invariant-mass
distributions alone is sometimes not sufficient to uniquely determine the spins
and coupling parameters. However, if the longer two-step cascade decay chain is
observed, and one can measure both the $\ell^+\ell^-$ and $j\ell^\pm$
invariant-mass distributions, it is possible to unambiguously discriminate
between all possible spin configuration with high significance. Furthermore, one
can independently constrain all coupling parameters and the mass of
the off-shell mediator $B$, up to an intrinsic two-fold ambiguity.

The results presented here are based on a parton-level analysis.
In a realistic experimental environment, the significance for the model
discrimination and the precision for the parameter determination
may be diluted by jet energy smearing and combinatorics, but the essential
features and main conclusions are not affected substantially by these effects.


\section*{Acknowledgements}

C.-Y.~C. acknowledges
support by The George E.~and Majorie S.~Pake Fellowship during part of this
project. Also, he is grateful for the hospitality of the
Theoretical Advanced Studies Institute (TASI 2011) at the University of
Colorado at Boulder, where part of this work was done.
The research of A.~F. is supported partially by the National Science Foundation
under grant PHY-0854782.


\section*{Appendix: Formulae for invariant-mass distributions}

Explicit expressions for the functions $f_i^{(\ell\ell)}$ and $f_i^{(j\ell)}$
are avaiable in \textsc{Mathematica} format at  {\tt
http://www.pitt.edu/\~{}afreitas/dec3.tgz}. Note that the expressions in this file
are not normalized, since in practice the normalization is best carried out
numerically as described in section~\ref{proc}. The results for
$f_i^{(\ell\ell)}$ are given as analytical formulae, while $f_i^{(j\ell)}$ are
presented in terms of one-dimensional integral representations of the form
\begin{equation}
f_i^{(j\ell)} = \int_{m_A^2}^{m_C^2[1-m_{q\ell^+}^2/(m_D^2-m_C^2)]}
dm_{A\ell^-}^2 \, F_i^{(j\ell)}.
\end{equation}


\end{document}